\documentclass[]{spiejour}

\usepackage{amsmath, amssymb, amsfonts}
\usepackage{graphicx, epsfig,wrapfig}

\usepackage{color}
\usepackage{subfigure}

\def\##1{\mathbf{#1}}
\def\=#1{\underline{\underline{#1}}}

\def\Jso{\mathbf{J}^{so}(\#r)}

\def\ucn{\#e_\rho}
\def\Rcn{R_{cn}}

\def\Imax{I_{max}}
\def\Iavg{I_{avg}}

\title{Scattering of the near field of an electric dipole by a single-wall carbon nanotube}

\author{Andrei M. Nemilentsau,\supscr{a} Gregory Ya. Slepyan,\supscr{a} Sergey A. Maksimenko,\supscr{a} Akhlesh Lakhtakia,\supscr{b} and Slava V. Rotkin\supscr{c}}

\affiliation{\supscr{a}Institute for Nuclear Problems, Belarus State University, 11 Bobruiskaya Street, Minsk 220030, Belarus\\
\linkable{andrei.nemilentsau@gmail.com} \\
\supscr{b}NanoMM--Nanoengineered Metamaterials Group, Department of Engineering Science and Mechanics, Pennsylvania State University, University Park, PA 16802, USA\\
\supscr{c}Department of Physics,  and Center for Advanced Materials and Nanotechnology,
Lehigh University, 16 Memorial Dr.~E., Bethlehem, PA 18015, USA\\
}

\begin{document}
\maketitle

\begin{abstract}
The use of carbon nanotubes as optical probes for  scanning near-field optical microscopy
requires an understanding of their near-field response. As a first
step in this direction, we investigated the lateral resolution of a carbon nanotube tip with respect to an ideal
electric dipole   representing an elementary detected object.
A Fredholm integral equation of the first kind was formulated for the surface electric current density induced on
a single-wall carbon nanotube (SWNT) by the electromagnetic field due to an arbitrarily
oriented electric dipole located outside the SWNT. The response of the SWNT to the
near field of a source electric dipole can be classified into two types, because
surface-wave propagation occurs with (i) low damping at frequencies less
than $\sim 200$-$250$~THz and (ii) high damping at higher frequencies. The
interaction between the source electric dipole and the SWNT depends critically
on their relative location and relative orientation, and shows evidence of the
geometrical resonances of the SWNT in the low-frequency regime. These resonances
disappear when the relaxation time of the SWNT is sufficiently low.   The far-field radiation intensity is much higher when the source electric dipole is placed near an edge of SWNT than at the centroid of the SWNT.  The use of an SWNT tip in scattering-type scanning near-field optical microscopy can deliver a resolution less than $\sim20$ nm. Moreover, our study shows that the relative orientation and distance between the SWNT and the nanoscale dipole source can be detected.

\end{abstract}


\section{Introduction}\label{intro}
Carbon nanotubes,  hollow cylindrical  rolls of graphene layers,
possess remarkable electronic properties \cite{Charlier}
as well as tremendous mechanical stability and strength \cite{Salvetat} that makes them
attractive for various applications in opto- and nano-electronics \cite{Avouris08}. One of the
focuses of current research
is on the responses of carbon nanotubes to an externally applied electromagnetic field \cite{Maksimenko,Rutherglen,Novotny},
because carbon nanotubes have prominent  absorption resonances in the infrared and visible regimes, the resonance
 frequencies reflecting the carbon nanotube lattice symmetry \cite{Reich}. Moreover, carbon nanotubes are strongly nonlinear optical structures
\cite{Margulis,Maksimenko,Wang_09} with an ultrafast optical response  \cite{Shen}  that
could be exploited for ultrafast
optical signal-processing devices.
Furthermore, as research advances, carbon nanotubes are no longer considered as infinitely long. The influence of the
finite length  through edge effects on the electromagnetic responses of carbon nanotubes is being investigated, particularly
for carbon nanotubes operating as receiving nanoantennas. Reports have been published
on isolated single-wall carbon nanotubes (SWNTs)
\cite{Hanson_04,Burke_06,Slepyan_06}, isolated multiwall carbon nanotubes
 \cite{Shuba_09}, bundles of carbon nanotubes \cite{Shuba_07}, and carbon nanotube arrays \cite{Wang_04a,Hao}.
Actual radio receivers utilizing an electromechanical
modulation of the field emission from a carbon nanotube antenna  have been successfully fabricated \cite{JWGZ07}.

Most research in this context is devoted to the response of a carbon nanotube to a uniform plane wave.  But a carbon nanotube's response to
the spatially nonuniform near field due to a closely located source ought to be substantially different from the planewave response of the
same carbon nanotube, depending on the relating location
and orientation of the carbon nanotube and the near-field's source, as is known to be true for other scatterers \cite{LIDH82a,LIDH82b}.
 This can also be inferred from a
theoretical analysis of the scattering of a plane wave jointly by a metallic nanosphere and an SWNT \cite{Hanson_07}.
More recently, the intensity spectrum of the thermal radiation from an SWNT in the near-field zone 
was found to have  additional resonant lines
than its analog when the SWNT is in the far-field zone
 \cite{Nemilentsau}.

 An understanding of the responses of carbon nanotubes in the near-field zone is
crucial for  scattering-type scanning near-field optical microscopy (sSNOM) and biomarking.
A sharp tip placed in
the proximity of the sample is used in sSNOM to scatter near fields induced on the sample surface by an external focused light beam. The scattered light
involves near-field zone interaction between the tip and the sample and maps the surface characteristics of the sample
in terms of a local refractive index and a local
absorption coefficient \cite{Novotny}. Thus, a resolution on the order of several nm could be obtained by using an SWNT as
a tip. Indeed, an resolution of 30 nm at 633-nm wavelength has already been reported
\cite{Hillenbrand}. In the rapidly developing field of the terahertz aperture-less near-field spectroscopy and
 microscopy \cite{Chan}, a
resolution of 100 nm has been obtained with a tungsten tip \cite{Chen,Thoma} and 30 nm
with a platinum tip \cite{Huber}, and a proposal for SWNTs as tips has already been made
 \cite{Slepyan_06}.

Another possibility for using carbon nanotubes for SNOM would
allow complementary spectroscopic characterization of the
photoluminescence (PL) properties of nanoscale objects. A higher near-field intensity due to the tip changes and amplifies the signal from a source that is either
a molecule or a man-made nano-object, with detection taking pave in the far-field zone \cite{Novotny2}. Scanning along the surface produces the PL map with
potentially subwavelength resolution being controlled by the apex radius
of the probe.

Nanostructures that combine such multiple  functionalities as biocompatibility,
fluorescent signalling, and drug storage and delivery, will advance cancer diagnostics and therapeutics.
Integration of carbon nanotubes with nanoscale
luminescent  materials such as quantum dots (QDs) appears promising \cite{Shi}. This integration
may be achieved by functionalizing carbon nanotubes with either DNA molecules
 \cite{Zhou} or carboxyl groups \cite{Shi,Wang_09a} to which QDs could be attached. Integration of carbon nanotubes and QDs
substantially affects the  luminescence properties of QDs, corroborating the
experimentally determined energy transfer between carbon nanotubes and QDs
 \cite{Wang_09a}.

The foregoing examples amply demonstrate that the near-field responses of carbon nanotubes require a comprehensive investigation.
As a first step in that direction, we examine the scattering of the near field of an oscillating, point electric dipole in the
proximity of an SWNT of finite length---to address the effect of PL amplification
and to understand the role of the SWNT antenna/near-field probe in the formation of the radiation pattern of a point PL source.
{The outline of the paper is as follows:} The boundary-value problem is formulated in Sec.~\ref{Ch:model},
and several typical  {profiles of the current induced on an SWNT} are presented and discussed in
Sec.~\ref{Ch:Numerical}. Our theory is intended to cover the coupling of a broad variety of luminescent nano-objects to an SWNT
near-field probe. As the energy transfer between a nano-object and an SWNT can be either resonant or non-resonant,
{depending on the excitation frequency,} 
illustrative examples are
presented in Sec.~\ref{Ch:Numerical}.  
{Section~\ref{Ch:Inside} is devoted to the special case of}
{an endohedral molecule inside the SWNT,} the molecule being modeled as
{a coparallel source electric dipole located on the SWNT axis. The resonant coupling
of an SWNT placed close to a source electric dipole is examined in 
Sec.~\ref{ch:plasmons}. Sec.~\ref{Ch:scattered} provides a detailed examination
of the scattered electric field. Examining the contour plots of the far-zone
field due to the dipole-SWNT system
in Sec.~\ref{Ch:Pattern}, we determine the spatial resolution that an SWNT tip can deliver in sSNOM.
Conclusions are provided in Sec.~\ref{Ch:Conclusion}. Gaussian units are used, and
a time dependence of $\exp(-i\omega t)$
is implicit with $t$ as time, $\omega$ as angular frequency, and $i=\sqrt{-1}$. Vectors are denoted in boldface;
unit vectors are denoted as $\#e_x$, etc.; and dyadics \cite{Chenbook} are double-underlined. 

\section{Boundary-value problem}
\label{Ch:model}

Suppose an SWNT of  length $L$ is exposed to the electromagnetic field radiated by a
current density $\Jso$, where $\mathbf{r}$ denotes the position vector.
The electric field  everywhere must satisfy the
nonhomogeneous vector Helmholtz equation
\begin{equation}
\left[\left(\nabla\times\=I\right)\cdot\left(\nabla\times\=I\right)-k^2\=I\right]\cdot
\mathbf{E}(\mathbf{r})=\frac{4\pi i \omega}{c^2}\Jso \,,
\label{Eq:Electric_field_sc}
\end{equation}
where
$k=\omega/c$ is the free-space wavenumber,  $c$ is the speed of
light in free space, and
$\=I$ denotes the identity dyadic.
Likewise, the magnetic field  everywhere must be a solution of the
related equation
\begin{equation}
\left[\left(\nabla\times\=I\right)\cdot\left(\nabla\times\=I\right)-k^2\=I\right]\cdot
\mathbf{H}(\mathbf{r})=\frac{4\pi }{c}\nabla\times\Jso \,.
\label{Eq:Magnetic_field_sc}
\end{equation}

If the cross-sectional radius of the SWNT is denoted by $R_{cn}$, the SWNT axis is aligned
parallel to the $z$ axis of a Cartesian coordinate system $(x,y,z)$, and the centroid of the SWNT is
designated as the origin of the coordinate system, any point on the surface of
the SWNT can be specified by
\begin{equation}
\mathbf{r}_{cn} = R_{cn}\left(\cos\phi\,\mathbf{e}_x+ \sin\phi\,\mathbf{e}_y\right)+z\mathbf{e}_z\,,\quad
\phi\in[0,2\pi)\,,\quad z\in[-0.5L,0.5L]\,,
\end{equation}
where $\#e_{x,y,z}$ are the Cartesian unit vectors.
Using the equivalent cylindrical coordinate system $(\rho,\phi,z)$, we have to enforce the satisfaction of the
following boundary conditions \cite{Slepyan_99,Slepyan_aeu}
 \begin{eqnarray}
\nonumber
&&\lim_{\delta\to0}\left\{\ucn\times
\left[\#H(\Rcn+\delta,\phi,z)-\#H(\Rcn-\delta,\phi,z) \right]\right\}
\\
\label{Eq:boundary1a}
&&\qquad\qquad=\frac{4\pi}{c} \#J^{eq}(z)\,,   \quad
z\in[-0.5L,0.5L]\,,\quad \phi\in[0,2\pi)\,,
\\
\nonumber
&&\lim_{\delta\to0}\left\{\ucn\times
\left[\#H(\Rcn+\delta,\phi,z)-\#H(\Rcn-\delta,\phi,z) \right]\right\}
\\
\label{Eq:boundary1b}
&&\qquad\qquad
=\#0\,,   \quad
z\notin[-0.5L,0.5L]\,,\quad \phi\in[0,2\pi)\,,
\\
\nonumber
&&\lim_{\delta\to0}\left\{\ucn\times
\left[\#E(\Rcn+\delta,\phi,z)-\#E(\Rcn-\delta,\phi,z) \right]\right\}
\\
\label{Eq:boundary1c}
&&\qquad\qquad=\#0\,,   \quad
z\in(-\infty,\infty)\,,\quad \phi\in[0,2\pi)\,.
\end{eqnarray}
Here, $\#J^{eq}(z)$, the surface
current density induced on the SWNT's surface $S$, is a measure of  the jump in the
tangential magnetic field across that surface.

The surface electric current density is assumed to be independent of
$\phi$ and purely axial: $\#J^{eq}(z)\equiv J^{eq}(z)\#e_z$. For this statement to be valid, the electric field radiated by the current density $\Jso$ needs to be homogeneous along the SWNT circumference. Thus the following restriction is imposed on the system under the consideration: $k\Rcn \ll 2\pi$. Based on the spatial variations of the electromagnetic field emitted by $\#J^{so}$ when
the scattering SWNT is absent, additional restrictions may emerge too.  Finally, it must
satisfy the edge conditions
\begin{equation}
\label{edge}
J^{eq}(\pm 0.5L)=0\,,
\end{equation}
which express the absence of concentrated charges on the
two edges of the SWNT.

Except in the source region, the electric field  can be represented as \cite{Chenbook,Strom}
\begin{equation}
\#E(\#r)= \#E^{inc}(\#r) + \#E^{sca}(\#r)\,,
\label{totalE}
\end{equation}
where the incident electric field
\begin{equation}
\#E^{inc}(\#r) =  \frac{i\omega}{c^2}\int_{V^{so}}\,d^3\#r^\prime\, \=G(\#r,\#r^\prime)\cdot\#J^{so}(\#r^\prime)
\label{Einc}
\end{equation}
is  due to the source current density that is confined to the region $V^{so}$ lying outside the SWNT,
and the scattered electric field  everywhere is given by
\begin{equation}
\#E^{sca}(\#r) =  \frac{i\omega}{c^2}\int_{S}\,d^2\#r^\prime\, \=G(\#r,\#r^\prime)\cdot\#J^{eq}(z^\prime).
\label{Esca}
\end{equation}
In these equations,
\begin{equation}
\=G(\#r,\#r^\prime) = \left(\=I + \frac{\nabla\nabla}{k^2}\right)
\,
\frac{\exp({ik\vert\#r-\#r^\prime\vert})}{\vert\#r-\#r^\prime\vert}
\end{equation}
is the dyadic free-space Green function. Equations can be similarly
written for the incident and the scattered magnetic fields, but are not needed.

An axial surface conductivity $\sigma_{zz}$ can be obtained
 to relate the jump in the tangential magnetic field
across the surface $\rho=\Rcn$ of the SWNT to the axial electric field \cite{Slepyan_99}; thus,
\begin{equation}
E_z(\#r_{cn})\equiv\#e_z\cdot\#E(\#r_{cn})=\frac{J^{eq}(z)}{\sigma_{zz}}\,, \quad z\in(-0.5L,0.5L)\,.
\label{Ohm}
\end{equation}
 This equation is quite general, except that it does not take into account the effects of spatial dispersion as well
 as excitonic effects; furthemore, we have neglected the contribution of the chiral conductivity $\sigma_{z\phi}$ to the axial current because that contribution is quite small in SWNTs \cite{Slepyan_98}.

Taking the dot product of both sides of Eq.~(\ref{Esca}) with $\#e_z$, and making use
of Eqs.~(\ref{totalE}) and (\ref{Ohm}) therein,
we obtain
\begin{equation} \label{Eq:One_dimension}
\frac{J^{eq}(z)}{\sigma_{zz}}-E_z^{inc}(\mathbf{r}_{cn})=\left(\frac{d^2}{d z^2} + k^2\right) \Pi(z)\,,
\quad z\in(-0.5L,0.5L)\,,
\end{equation}
where the scalar Hertz potential
\begin{equation} \label{Eq:Hertz_sol}
\Pi(z)=\frac{i R_{cn}}{\omega } \int_{-0.5L}^{0.5L} dz^\prime\, J^{eq}(z^\prime)\int_{0}^{2\pi} d\phi^\prime \,\frac
{\exp\left[ik\sqrt{(z-z^\prime)^2 + 4 R_{cn}^2 \,\sin^2(\phi^\prime/2)}\right]}
{\sqrt{(z-z^\prime)^2 + 4 R_{cn}^2 \,\sin^2(\phi^\prime/2)}}\,.
\end{equation}
According to our assumptions, $E_z^{{inc}}(\mathbf{r}_{cn})$ changes so slowly along the SWNT circumference that it
can be assumed to
 depend on the axial coordinate only; i.e., $E_z^{{inc}}(\mathbf{r}_{cn}) \simeq
 E_z^{{inc}}(0,0,z_{cn})$.
Thereby, Eq.~(\ref{Eq:One_dimension}) becomes an integrodifferential equation for $J^{eq}(z)$
 with the formal solution
\begin{equation} \label{Eq:Sol_one_dim}
\Pi(z)=C_1 e^{-i k z} + C_2 e^{i k z} + \frac{1}{2ik}\int_{-0.5L}^{0.5L}\,dz^\prime\,\left\{e^{ik\vert z - z^\prime\vert}
\left[\frac{J^{eq}(z^\prime)}{\sigma_{zz}}
-E_z^{{inc}}(0,0,z^\prime)\right]\right\}\,,
\end{equation}
where the constants $C_{1,2}$ have to be determined eventually using
the edge conditions (\ref{edge}).

Equating the right sides of Eqs.~(\ref{Eq:Hertz_sol}) and (\ref{Eq:Sol_one_dim}), we obtain
\begin{equation} \label{Eq:Integr_eq}
 C_1 e^{-i k z} + C_2 e^{i k z} +
\int\limits_{-0.5L}^{0.5L}\,dz^\prime\,  {\mathcal K}(z-z') J^{eq}(z^\prime)
= \frac{1}{2 i k} \int\limits_{-0.5L}^{0.5L}\,
dz^\prime\, e^{ik\vert z - z^\prime\vert}\,E_z^{{inc}}(0,0,z^\prime) \,,
\end{equation}
with the kernel
\begin{equation} \label{Eq:Integral_kernel}
 {\mathcal K}(z) = \frac{e^{i k \vert z\vert}}{2 i k \sigma_{zz}} + \frac{R_{cn}}{i\omega} \int_0^{2\pi}
 d\phi  \,\frac
{\exp\left[ik\sqrt{z^2 + 4 R_{cn}^2 \,\sin^2(\phi/2)}\right]}
{\sqrt{z^2 + 4 R_{cn}^2 \,\sin^2(\phi/2)}}
\,.
\end{equation}
Equation (\ref{Eq:Integr_eq}) is a Fredholm integral equation of the first kind with $J^{eq}(z)$
as the unknown function to be determined \cite{Arfken}. It has to be solved numerically for specified
$E_z^{inc}(0,0,z)$, as in a predecessor paper \cite{Slepyan_06}. Once $J^{eq}(z)$ has been determined
for all $z\in(-0.5L,0.5L)$, the scattered electric field can be calculated at any location using
Eqs.~(\ref{edge}) and (\ref{Esca}).

We chose the source of the near field to be a
point electric dipole of moment $\mathbf{p}_0$ located at $\#r_s$ outside
the SWNT, as shown in Fig.~\ref{Fig:system4}; accordingly,
\begin{equation} \label{Eq:delta-source}
 \Jso= - i \omega \mathbf{p}_0 \delta(\mathbf{r}-\mathbf{r}_s).
\end{equation}
Substituting  Eq.~(\ref{Eq:delta-source}) on the right side of Eq.~(\ref{Einc}),
we obtained
\begin{equation} \label{Eq:Field_inc1}
\mathbf{E}^{{inc}}(\mathbf{r}) = k^2 \, \=G(\mathbf{r},\mathbf{r}_s) \cdot \mathbf{p}_0.
\end{equation}
The electromagnetic field due to an electric dipole contains
terms that vary as $\vert\#r-\#r_s\vert^{-3}$, $\vert\#r-\#r_s\vert^{-2}$,  and $\vert\#r-\#r_s\vert^{-1}$,
with the first term dominant in the near-field zone and the third term dominant in the far-field zone \cite{Chenbook}.
There are certain restrictions on $\#p_0$ and $\#r_s$ to ensure that the $z$-directed component
of the incident electric field varies very little along the circumference of the SWNT for any $z\in(-L/2,L/2)$.

In the foregoing equations, we have implicitly assumed that
the {source} incorporates interaction with the SWNT, and is
therefore {in a steady state} {\em in the presence} of the SWNT. As such, it already includes possible
renormalization of the dipole frequency (and lifetime). Such quantities can be computed knowing
the ones in the absence of the SWNT, {but that substantial exercise will be taken up in the future.}

\begin{figure}[ht!]
\begin{center}
\includegraphics[width=6cm]{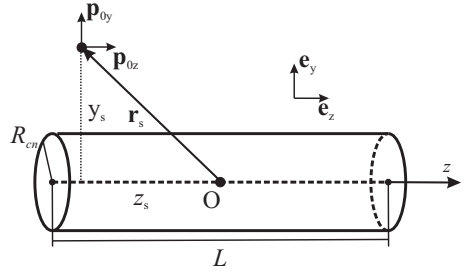}\\[2mm]
\small{}
\caption{Schematic for the scattering by an SWNT of the electromagnetic field
due to an electric dipole of moment $\#p_0$. }\label{Fig:system4}
\end{center}
\end{figure}

\section{Typical Profiles of Induced Surface Current}
\label{Ch:Numerical}

Calculations were performed for two types of metallic  {and one type of semiconductor zigzag SWNTs}---specified by the
dual
indexes $(14,0)$, $(15,0)$ and $(18,0)$, as is commonplace \cite{Maksimenko,Reich}---each of length $L = 1$~$\mu$m.
The axial surface conductivity $\sigma_{zz}$ was computed using the relaxation time
$\tau = 3\times 10^{-12}$~s  (except for those in Sec.~\ref{ch:plasmons}) and the overlap integral
$\gamma_0\approx 2.7$~eV, as shown elsewhere \cite{Slepyan_99}.

\begin{figure}[h]
\centering%
\subfigure[$\#p_0=10^{-20}\,{\#e_z}$~esu, $\#r_s= 10\,\#e_y$~nm%
\label{Fig:currentz_centr}]%
{\includegraphics[width=10cm]{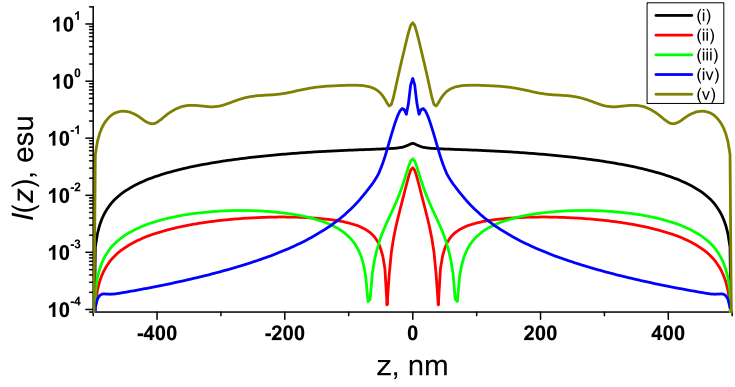}} %
\subfigure[$\#p_0=10^{-20}\,{\#e_y}$~esu, $\#r_s= 10\,\#e_y$~nm%
\label{Fig:currenty_centr}]%
{\includegraphics[width=10cm]{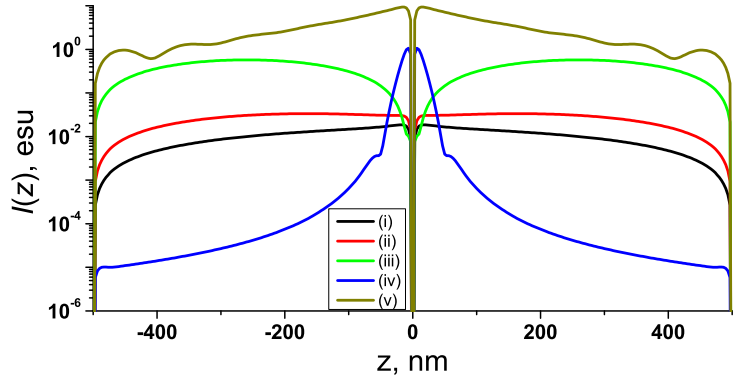}}%
\caption{\label{Fig:current_centr}
Surface current $I(z)$ induced along the axis of a $(15,0)$ SWNT
of length $L=1$~$\mu$m in response to a
source electric dipole $\#p_0$ located at $\#r_s$ in the central transverse plane
of the SWNT. The electric dipole is oriented either (a) parallel or (b) normal to the axis of the SWNT.
}
\end{figure}

Two orientations of the electric dipole were considered: parallel to the
SWNT axis and normal to the SWNT axis. 
 The magnitude $p_0$ of the electric dipole moment  was assumed  equal to
 $0.01$~D $=10^{-20}$~esu, while
 the location $\#r_s$ was varied.  {We illustrate the frequency dependence of the electromagnetic coupling between the
 source and the SWNT in this section by computing profiles of the induced
 current at the following
 four different frequencies}:

 \begin{itemize}

 \item[(i)] $\omega/2\pi= 2.6$~THz, the first geometrical-resonance frequency for surface-wave
 propagation, defined by the condition $(h^2-k^2) L^2 = \pi^2$ \cite{Slepyan_06}, where $h$ is the
 guide wavenumber \cite[Eq. (58)]{Slepyan_99};

  \item[(ii)] $\omega/2\pi= 4.0$~THz, an off-resonance frequency;

  \item[(iii)] $\omega/2\pi= 5.2$~THz, the second geometrical-resonance frequency for surface-wave
 propagation, defined by the condition $(h^2-k^2) L^2 = 4\pi^2$ \cite{Slepyan_06,Slepyan_99}; and

\item[(iv)] $\omega/2\pi= 500$~THz, the frequency of interband transitions between
Van Hove singularities \cite{VanHove,Bassani}.

 \end{itemize}
 In addition, for the $(18,0)$ SWNT we chose
 \begin{itemize}
 \item[(v)] $\omega/2\pi= 1310$~THz, a plasmon resonance frequency
 defined by the condition $\hbar \omega =2 \gamma_0$ \cite{Slepyan_06,FPEH05}.
 \end{itemize}
 The smallest free-space wavelength $\lambda=2\pi/k$ among these five cases is 229~nm. As
 the cross-sectional radius   of the $(15,0)$ SWNT is
 $0.587$~nm and that of the $(18,0)$ SWNT is $0.705$~nm ,
both  satisfy the {condition} $kR_{cn}\ll 1$.

\subsection{Coupling to the near field of a source electric dipole}

\begin{figure}[h]
\centering%
\includegraphics[width=10cm]{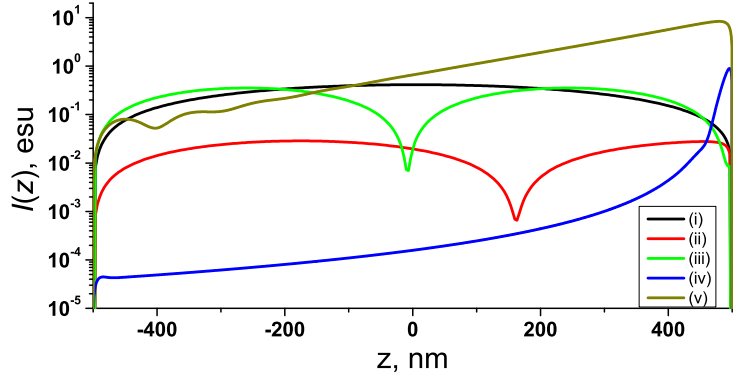}
\caption{\label{Fig:current_edge} Same as Fig.~\ref{Fig:currentz_centr}, except that the  source electric dipole is located at $\#r_s$ along the axis
of the $(15,0)$ SWNT ($\#p_0=10^{-20}\,{\#e_z}$~esu and $\#r_s= 510\,\#e_z$~nm). }
\end{figure}

The calculated profiles of the electric current $I(z) = 2\pi \Rcn \vert{J^{eq}(z)}\vert$ induced on the SWNT surface by the electric field (\ref{Eq:Field_inc1})
due to a source electric dipole are presented in
\begin{itemize}
\item Fig. \ref{Fig:currentz_centr}  for $\#r_s=10\,\#e_y$~nm and $\#p_0=10^{-20}\,\#e_z$~esu,
\item Fig. \ref{Fig:currenty_centr}  for $\#r_s=10\,\#e_y$~nm and $\#p_0=10^{-20}\,\#e_y$~esu, and
\item Fig. \ref{Fig:current_edge}  for $\#r_s=510\,\#e_z$~nm and $\#p_0=10^{-20}\,\#e_z$~esu.
\end{itemize}
The remaining case---$\#r_s=510\,\#e_z$~nm and $\#p_0=10^{-20}\,\#e_y$~esu---to complete
a quartet turned out to be
trivial for the following reason: When the electric dipole is situated on the SWNT axis and is directed
normal to that axis, the $z$-directed component of the incident electric field changes sign along
any cross-sectional diameter of the SWNT. Thus, that component of the incident electric field cannot be
assumed to be invariant across the circumference of the SWNT. However,
 as that component in one half of the SWNT bisected by the meridional
 plane to which the electric dipole is tangential is opposite in sign to  that component in the other
 half, and {consistently
 with the fully symmetric 
 one-dimensional approximation allowed by the condition} $kR_{cn}\ll 1$, we can assume that the coupling between the source electric dipole
 and the SWNT is negligibly small, at least to the first approximation.

In all other cases (Figs. \ref{Fig:current_centr} and \ref{Fig:current_edge}) the
source electric dipole is located
10~nm from the closest point on the SWNT, which distance is less than $\lambda/20$ even at the largest frequency considered. Therefore, not surprisingly, the closest part of the SWNT is tightly coupled to the electric dipole
and sustains a very large surface current density. Apart from that feature, which is common to all curves in
Figs. \ref{Fig:current_centr} and \ref{Fig:current_edge}, {\it two} different regimes of the SWNT's response to the near field can be identified.

The first regime
is the low-frequency (terahertz) regime for 
metallic SWNTs---represented by cases (i)--(iii)---wherein the SWNTs support surface-wave propagation with very low damping. In this regime the distribution of the surface current density is quite uniform
along the length of the SWNT, subject, of course, to the edge conditions (\ref{edge}).

The second regime of the SWNT's response to the near field is represented by cases (iv) and (v).  These response
characteristics  are inherent
\begin{itemize}
\item to metallic SWNTs in the high-frequency regime, where the contribution to $\sigma_{zz}$ from the interband electronic transitions becomes essential, and
\item to semiconducting SWNTs in the whole frequency range, as we show next.
\end{itemize}
Surface-wave propagation on the SWNT still occurs  in the second regime, but with high damping [case (v)] and very high damping [case (iv)]. Accordingly, the surface-current-density profile
is strongly peaked [case (v)] and very strongly peaked [case~(iv)] closest to the electric dipole.

The surface current profile along the SWNT axis does not depend directly on the free-space wavelength $\lambda$, as the near  field of dipole is strongly localized in the vicinity of the dipole. In order to
confirm this issue, the current induced in the (14,0) SWNT by the electric dipole oscillating at the frequency 2.6~THz---which
is the frequency of the first geometrical resonance in the (15,0) SWNT---is presented in Fig.~\ref{Fig:currentz_sem}. The dipole orientation and location are  the same as for Fig. \ref{Fig:currentz_centr}: $\#p_0=10^{-20}\,{\#e_z}$~esu and $\#r_s= 10\,\#e_y$~nm. The current distribution in the $(14,0)$ SWNT in Fig. \ref{Fig:currentz_sem} strongly resembles the current distribution in the $(15,0)$ SWNT at 500~THz---see Fig.~\ref{Fig:currentz_centr}---though the wavelength is 200 times smaller in the second case. The only difference is the maximum value of  the induced current $I_{max}$ that is much higher for
the $(15,0)$ SWNT at  500~THz than for the $(14,0)$ SWNT at  2.6~THz.

\begin{figure}[ht!]
\begin{center}
\includegraphics[width=10cm]{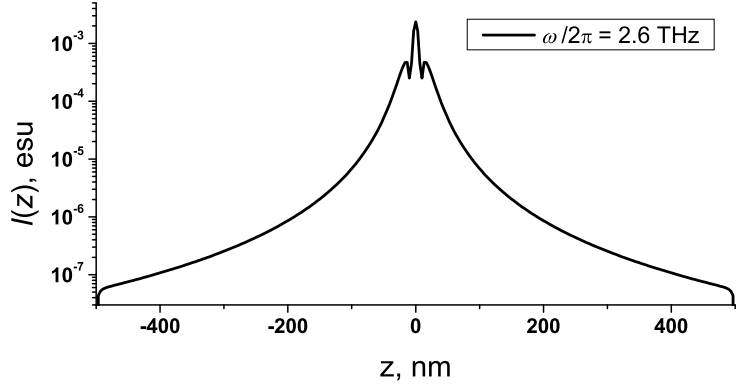}\\[2mm]
\small{}
\caption{Same as Fig. \ref{Fig:currentz_centr}, but for a $(14,0)$ SWNT.}
 \label{Fig:currentz_sem}
\end{center}
\end{figure}

The orientation of the electric dipole with respect to the SWNT axis has a significant effect on the profile of the induced current.
Let us first look at Fig.~\ref{Fig:current_centr}, for the source electric
dipole located in the central transverse plane of the SWNT.
At any point on the SWNT
axis, the incident electric field
has only $y$- and $z$-directed components for $z\in\left[-0.5L,0.5L\right]$. In the central part of the SWNT,
the incident electric field is primarily axial ($z$-directed) when the electric dipole is parallel,
but primarily normal ($y$-directed) when the electric dipole is normal, to the SWNT axis. Therefore,
following Eq.~(\ref{Ohm}), at the center of the SWNT
the induced surface current  is not null-valued in Fig.~\ref{Fig:currentz_centr}
but is null-valued in Fig.~\ref{Fig:currenty_centr}. Next, when
the electric dipole is located on the SWNT axis, as for Fig.~\ref{Fig:current_edge}, the difference is even more prominent
as a dipole oriented normal to the SWNT axis is not coupled to the SWNT   at all.

\subsection{Coupling to the far field of a source electric dipole }
\label{Ch:far-dipole}

Thus far we have considered the SWNT located in the near-field zone
of the source electric dipole.
It is instructive to consider also the case when the SWNT is located in the
far-field zone. Although this situation is not helpful to our goal 
of evaluating the possibility of SNOM applications, it gives additional insight on the interplay of the geometrical and frequency dependencies of the
coupling between the source field and the electronic system of the SWNT.


\begin{figure}[ht!]
\begin{center}
\includegraphics[width=10cm]{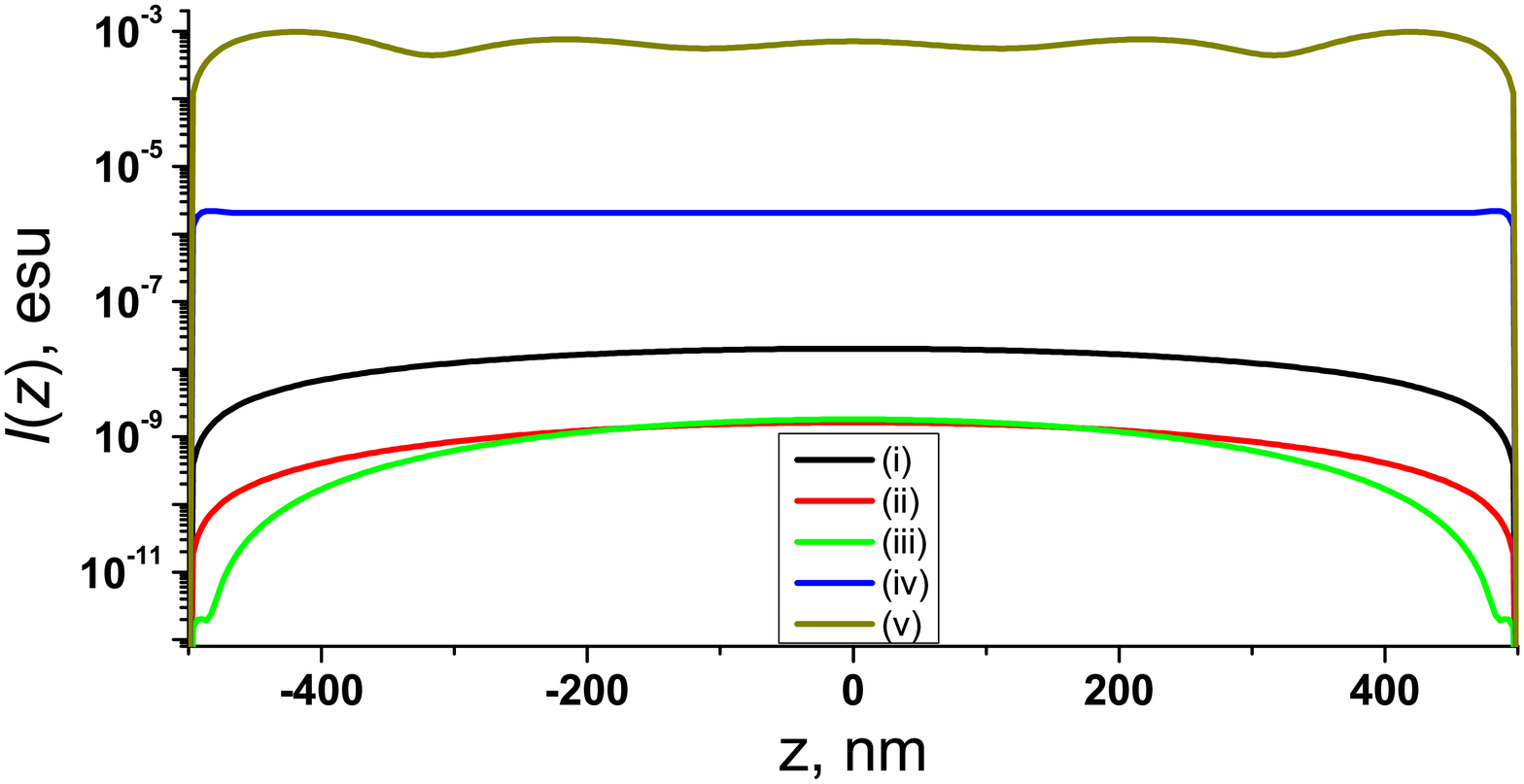}\\[2mm]
\small{}
\caption{Same as Fig. \ref{Fig:currentz_centr}, except for   $\#r_s = 100\,\#e_y$~$\mu$m.}
 \label{Fig:currentz_centr_far}
\end{center}
\end{figure}

The  profiles of the induced surface current computed for $\#r_s = 100\,\#e_y$~$\mu$m and $\#p_0=10^{-20}\,{\#e_z}$~esu
are presented in Fig.~\ref{Fig:currentz_centr_far}.  Two
response regimes are evident in this figure, just as
in Fig.~\ref{Fig:currentz_centr}: one regime covers cases (i)--(iii), and the other spans cases (iv) and (v).
The magnitude of the induced current is much smaller when $\#r_s = 100\,\#e_y$~$\mu$m than when
$\#r_s = 10\,\#e_y$~nm, because the incident electric field is much weaker in magnitude at any point
on the SWNT when the SWNT is irradiated by the far field.
The
profile of the induced current tends to be more uniform for $\#r_s = 100\,\#e_y$~$\mu$m than for
$\#r_s = 10\,\#e_y$~nm, because every point on the SWNT lies in the far-field zone of the source
electric dipole in the former scenario.

The
oscillatory profile of the induced surface current in Fig.~\ref{Fig:currentz_centr_far} for case (v) is due
to the fact that the wavelength of the surface wave   is $\sim200$~nm. In order to
explain this issue, let us recall that a surface wave with $\exp(ihz)$ dependence on $z$ can travel
on the SWNT, per \cite[Eq. (57)]{Slepyan_99}. The value of $h$ works out equal to
$h = 3.1 \times 10^{5} + i6.4 \times 10^{4}$~cm$^{-1}$, by virtue of
\cite[Eq. (58)]{Slepyan_99}, and the surface wave therefore has wavelength $2\pi/{\rm Re}(h)\approx
200$~nm.

\subsection{Coupling to a source electric dipole inside the SWNT}
\label{Ch:Inside}

The formulation presented in Sec.~\ref{Ch:Numerical} can be easily modified to to accommodate
the location of the source electric dipole inside the SWNT, so long as the $z$-directed component
of the incident electric field depends on the axial coordinate only. This situation mimics
 the special case of an endohedral molecule inside the SWNT, the molecule being modeled as a coparallel source electric dipole located on the SWNT axis.

\begin{figure}[!htb]
\begin{center}
\includegraphics[width=10cm]{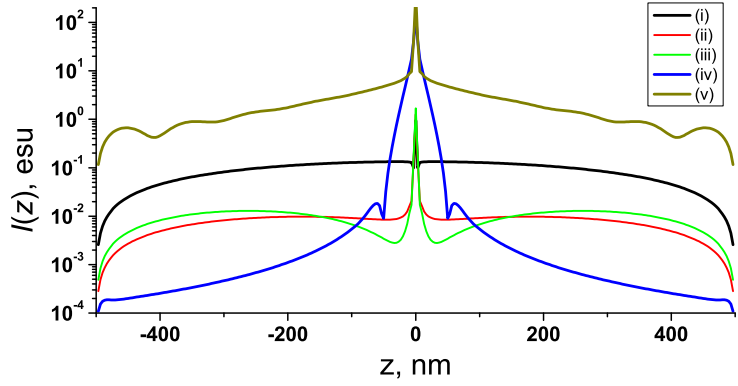}\\[2mm]
\small{}
\caption{Same as Fig. \ref{Fig:currentz_centr}, except for the source electric
dipole placed at the centroid of the $(15,0)$ SWNT  ($\#r_s = \#0$ nm).}
 \label{Fig:current_inside}
\end{center}
\end{figure}

Accordingly, after setting $\#p_0\parallel\#e_z$, $\#r_s=z_s\#e_z$, and $z_s\in (-0.5L,0.5L)$, we have to modify
Eq.~(\ref{Eq:Sol_one_dim}) to
\begin{eqnarray}
\nonumber
\Pi(z)&=&C_1 e^{-i k z} + C_2 e^{i k z}
+  \frac{1}{2ik}\int_{-0.5L}^{0.5L}\,dz^\prime\,\Bigg\{e^{ik\vert z - z^\prime\vert}
\\[5pt]
&&\qquad
\left[\frac{J^{eq}(z^\prime)}{\sigma_{zz}}
-E_z^{{inc}}\left(R_{cn}\cos\phi,R_{cn}\sin\phi,z^\prime\right)\right]\Bigg\}\,,
\label{Eq:Sol_one_dim-in}
\end{eqnarray}
with the tacit understanding that $E_z^{{inc}}\left(R_{cn}\cos\phi,R_{cn}\sin\phi,z^\prime\right)$
is independent of $\phi$.

The profiles of the surface current induced in the SWNT by a source electric
dipole placed at the centroid of the SWNT ($\#r_s = \#0$ nm) are presented in Fig.~\ref{Fig:current_inside}. These profiles  resemble the ones in Fig. \ref{Fig:currentz_centr} for the electric dipole placed at $\#r_s = 10\,\#e_y$ nm. However, we observe much more prominent rise in the induced surface current in the parts of the SWNT close to the source
electric dipole in the former case due to the higher strength of the incident electric field at  the SWNT surface.

\section{Resonant coupling in the near-field zone}
\label{ch:plasmons}

The edge conditions affect not only the profile  but also the  magnitude of the induced surface current, the latter being captured by the average induced surface current
\begin{equation} \label{Eq:AverageCurrent}
I_{avg} = \frac{1}{L} \int_{-L/2}^{L/2} I(z) dz.
\end{equation}

\begin{figure}[htb]
\begin{center}
\includegraphics[width=10cm]{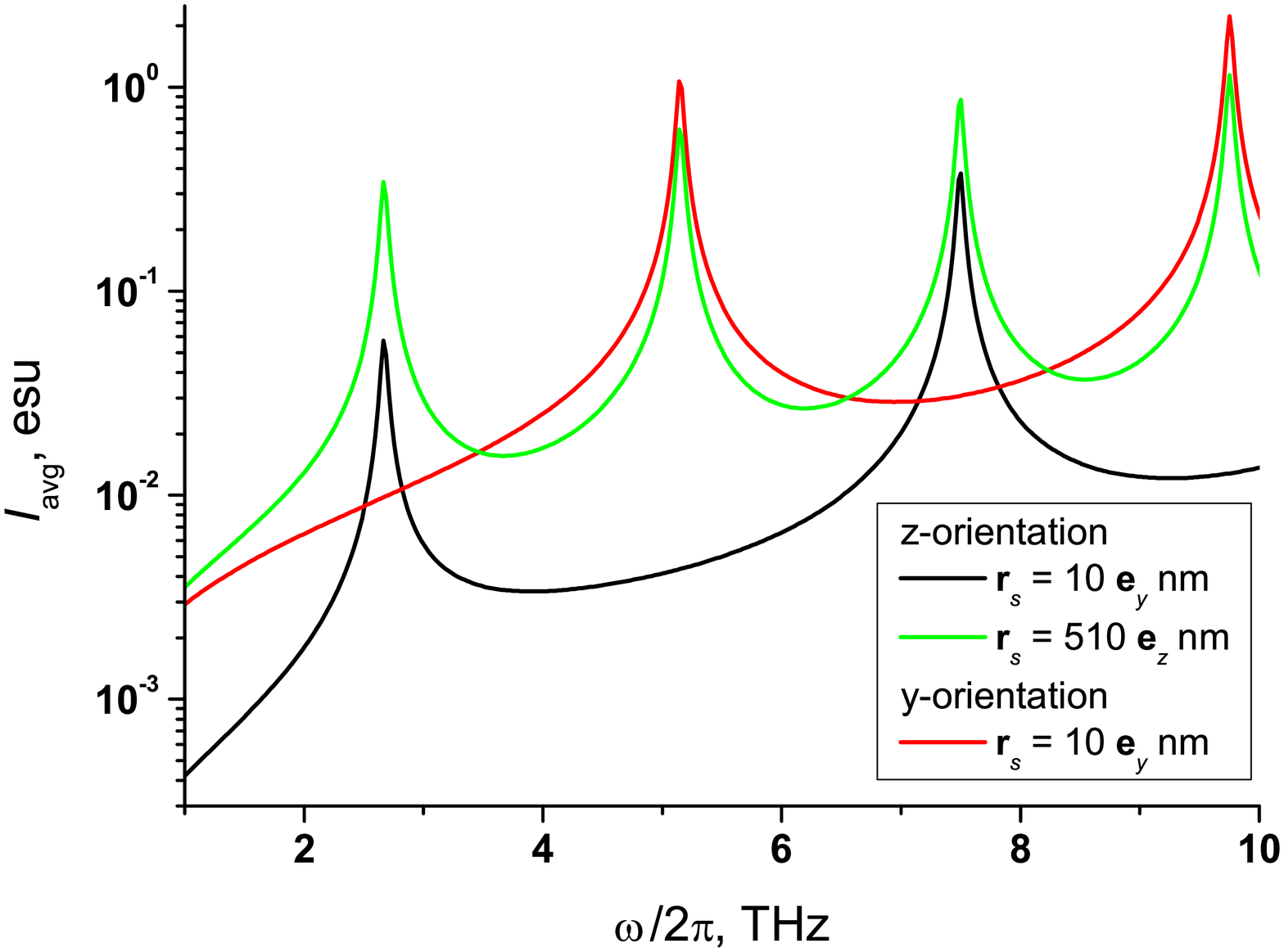}\\[2mm]
\small{}
\caption{Frequency dependence of the average surface current, defined
by  Eq.~(\ref{Eq:AverageCurrent}), induced in a $(15,0)$ SWNT due to a source electric dipole for three different  configurations: $\mathbf{p}_0 = 10^{-20}\, \mathbf{e}_z$ esu and $\mathbf{r}_s = 10\, \mathbf{e}_y$ nm (black line); $\mathbf{p}_0 = 10^{-20}\, \mathbf{e}_z$ esu and $\mathbf{r}_s = 510 \,\mathbf{e}_z$ nm (green line); $\mathbf{p}_0 = 10^{-20} \,\mathbf{e}_y$~esu and $\mathbf{r}_s = 10\, \mathbf{e}_y$~nm (red line).
}
 \label{Fig:AverageCurrent}
\end{center}
\end{figure}

The computed spectra of $\Iavg$ presented in Fig. \ref{Fig:AverageCurrent} contain several resonances at the frequencies defined by the condition $h L \approx \pi s$, where $s$ depends on the location and the orientation of the source electric dipole. When the electric dipole is situated near an edge of the SWNT, $s$ is an integer; when  the 
electric dipole is located equally distant from both edges of the SWNT,
$s$ is either even or odd, depending on the orientation of the electric dipole.

\begin{figure}[htb]
\begin{center}
\includegraphics[width=10cm]{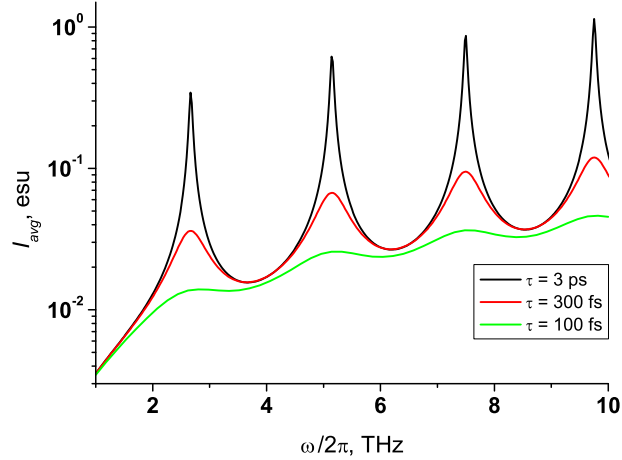}\\[2mm]
\small{}
\caption{Frequency dependence of $\Iavg$ in a $(15,0)$ SWNT due to a source electric dipole of moment $\mathbf{p}_0 = 10^{-20} \,\mathbf{e}_z$~esu located at $\mathbf{r}_s =510 \,\mathbf{e}_z$ for three different values
of the relaxation time $\tau$.
}
 \label{Fig:AverageCurrentRelaxation}
\end{center}
\end{figure}

Even though
the resonant amplitudes of $\Iavg$ in Fig. \ref{Fig:AverageCurrent} depend on the location and the orientation of
the source electric dipole, the 
full width at half maximum (FWHM) of each resonance is primarily defined not by the  source characteristics but by the attenuation of surface waves  in the SWNT. This attenuation
emerges in our model   through the relaxation time $\tau$ that
affects the axial surface conductivity $\sigma_{zz}$ of the
SWNT. In order to demonstrate the effect of
$\tau$, we present the spectra of $I_{avg}$ for different values of   $\tau$ in 
Fig.~\ref{Fig:AverageCurrentRelaxation}. A decrease in $\tau$ weakens the resonances but does not drastically affect
the off-resonance values of $\Iavg$.  The dissipation of the surface waves becomes so high  for $\tau< 100$~fs that the  resonances disappear.

It may seem from Figs.~\ref{Fig:current_centr} and \ref{Fig:current_edge}
that the maximum value $\Imax$ of the induced surface current $I(z)$  grows with the frequency and, in general, is much higher at optical frequencies [cases (iv) and (v)] than at terahertz
frequencies [cases (i)--(iii)]. However, this is not so. The spectrum of
$\Imax$ is presented in Fig. \ref{Fig:MaxCurrent}(a) for a $(15,0)$ SWNT when
a source electric dipole  $\mathbf{p} = 10^{-20}\, \mathbf{e}_z$~esu is located
at $\mathbf{r}_s = 10\, \mathbf{e}_y$~nm. The spectrum of the magnitude of the axial
surface
conductivity of the same SWNT   is presented in Fig.~\ref{Fig:MaxCurrent}(b), wherein we can identify two different spectral regimes: a Drude regime (frequencies $< 280$~THz) and a regime of interband transitions (frequency $ > 280$~THz). These two regimes are separated by a strong dip  (around $280$~THz). The spectral characteristics of $I_{max}$ are completely different in these two spectral regimes. In the low-frequency regime, $I_{max}$ varies
non-monotonically with frequency even though $\vert\sigma_{zz}\vert$ does.
 In particular we observe a number of resonant lines in the spectrum of $I_{max}$ that are the geometrical resonances of the surface plasmons---the same
 as in Fig.~\ref{Fig:AverageCurrent}. In the low-frequency regime ($<10$~THz)
 except for exactly at the resonance frequencies, $I_{max}$   is smaller than at the optical resonance frequencies.    In the high-frequency regime, the frequency dependence of $I_{max}$  follows the frequency dependence of $\vert\sigma_{zz}\vert$;
in particular, two poles
due to interband transitions are clearly seen in both spectra. 

\begin{figure}[ht!]
\begin{center}
\includegraphics[width=\textwidth]{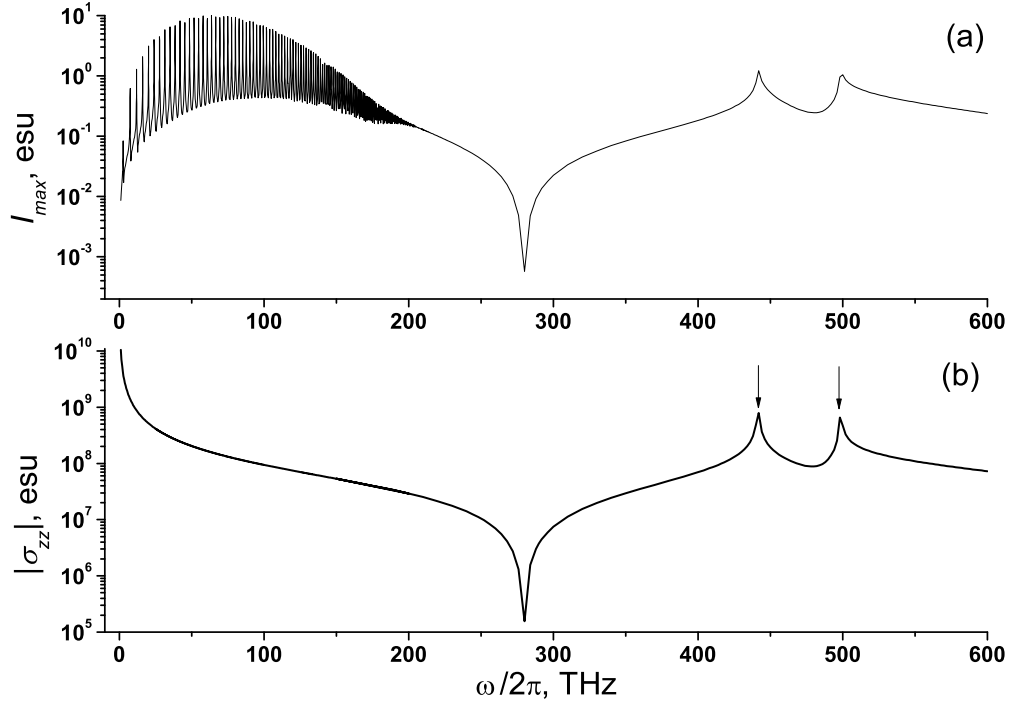}\\[2mm]
\small{}
\caption{(a) Spectrum of the maximum value $\Imax$ of the surface current $I(z)$ induced in a $(15,0)$
SWNT by an electric dipole of moment $\mathbf{p}_0 = 10^{-20} \,\mathbf{e}_z$~esu
located at $\mathbf{r}_s = 10\, \mathbf{e}_y$~nm. (b) Spectrum of
magnitude of the axial surface conductivity $\vert\sigma_{zz}\vert$ of a
$(15,0)$ SWNT. The resonances due to interband transitions   are indicated by arrows. }
 \label{Fig:MaxCurrent}
\end{center}
\end{figure}

To explain the spectral dependence of $I_{max}$, we should take into account that,
according to Eq.~(\ref{Eq:One_dimension}), the 
induced surface current density in the SWNT arises from the superposition of two electric fields;
i.e., 
\begin{equation} 
{J^{eq}(z)} =
{\sigma_{zz}}\left[E_z^{inc}(\mathbf{r}_{cn})+\left(\frac{d^2}{d z^2} + k^2\right) \Pi(z)\right]\,,
\quad z\in(-0.5L,0.5L)\,,
\end{equation}
where the second term within
the square brackets on the right side is the electric field of a surface wave and $\Pi(z)$ is defined by Eq.~(\ref{Eq:Hertz_sol}). The nonmonotonic dependence of $\Imax$ on
$\sigma_{zz}$ in the low-frequency regime in Fig. \ref{Fig:MaxCurrent}
 is because the electric field of the surface wave depends on the axial surface conductivity too. 
 
 Let us recall that 
 \begin{equation}
 E^{sca}_z(\#r_{cn}) = \left(\frac{d^2}{d z^2} + k^2\right) \Pi(z)\,,
 \quad z\in(-0.5L,0.5L)\,.
 \end{equation}
 Plotted in Fig.~\ref{Fig:StrengthFrequency}  are   the $z$-directed components of the incident and the scattered electric fields on the surface of
 a $(15,0)$ SWNT  illuminated by an electric dipole of moment
 $\mathbf{p}_0 = 10^{-20} \,\mathbf{e}_z$~esu located at $\mathbf{r}_s = 10\, \mathbf{e}_y$ nm. 
 At the frequency ($=2.6$~THz) of the first geometric resonance in the portion of
 the  SWNT close to the source electric dipole,  the $z$-directed components
 of the incident and the scattered electric fields  are of similar
 magnitude; see Fig.~\ref{Fig:StrengthFrequency}(a). However, as
 the real parts of the $z$-directed components
 of the two fields differ in sign---see Fig.~\ref{Fig:StrengthFrequency}(b)---the
 total electric field on the surface of the SWNT is small, leading to small
 $\Imax$ despite the high magnitude of the axial surface conductivity. As the
 frequency rises, the scattered electric field on the surface of the SWNT rises (especially at the geometrical resonances frequencies),
 as shown in Fig.~\ref{Fig:StrengthFrequency}(c),
 and leads to higher $\Imax$ despite a lowering of the surface axial conductivity.
  In the high-frequency regime, the magnitude of the axial surface conductivity is rather low (except for the interband transitions) and the surface wave  is strongly damped, the scattered electric field is generally much weaker in strength than the incident electric field,
  as shown in  Fig. \ref{Fig:StrengthFrequency}(d); hence,  $J^{eq}(z) \approx \sigma_{zz} E_z^{inc}(\mathbf{r}_{cn})$ does not depend on the scattered field.  

\begin{figure}[ht!]
\begin{center}
\includegraphics[width=\textwidth]{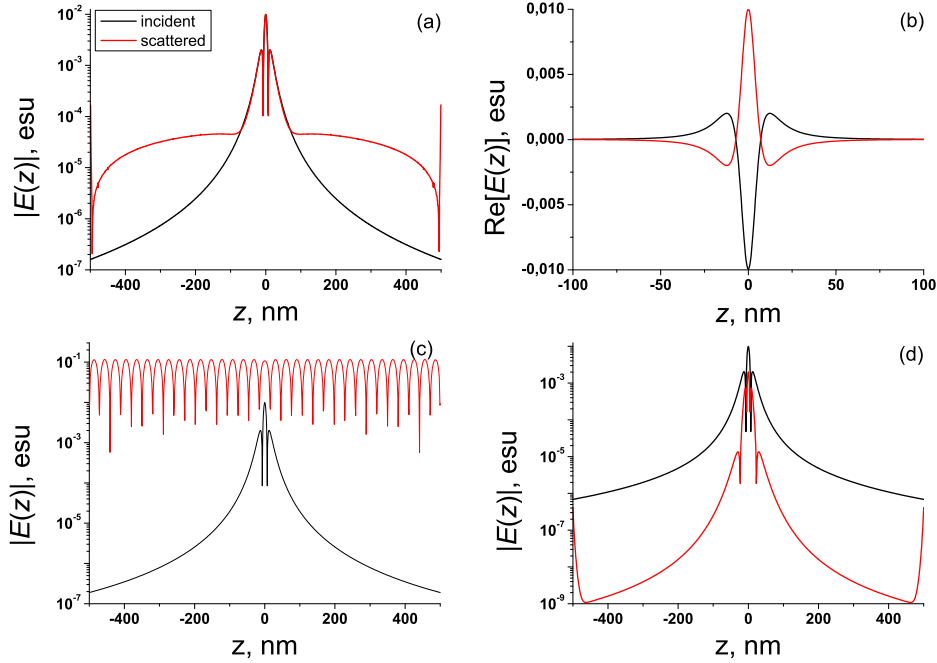}\\[2mm]
\small{}
\caption{(a) $\vert E_z^{inc}(z)\vert$ and $\vert E_z^{sca}(z)\vert$ at 2.6~THz,
(b) ${\rm Re}\left[E_z^{inc}(z)\right]$ and ${\rm Re}\left[E_z^{sca}(z)\right]$
at 2.6~THz, (c)  $\vert E_z^{inc}(z)\vert$ and $\vert E_z^{sca}(z)\vert$  at 60.9~THz, and
(d) $\vert E_z^{inc}(z)\vert$ and $\vert E_z^{sca}(z)\vert$ at 400~THz,
on the surface of a $(15,0)$ SWNT illuminated by
a source electric dipole  of moment $\mathbf{p}_0 = 10^{-20}\, \mathbf{e}_z$~esu is located
at $\mathbf{r}_s = 10\, \mathbf{e}_y$~nm.
 }
 \label{Fig:StrengthFrequency}
\end{center}
\end{figure}

\section{Scattered electric field}
\label{Ch:scattered}

\subsection{Scattered electric field near the SWNT}
\label{Ch:scatered_near_field}


\begin{figure}[h]
\centering%
\subfigure[$\#p_0=10^{-20}\,{\#e_z}$~esu, $\#r_s= 10\,\#e_y$~nm%
\label{Fig:efieldz_centr}]%
{\includegraphics[width=10cm]{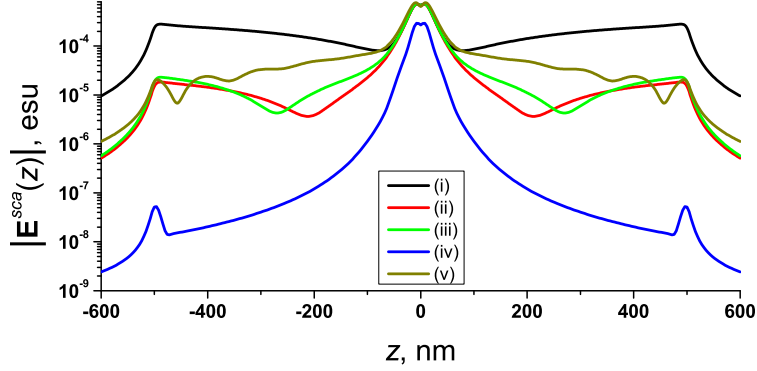}} %
\subfigure[$\#p_0=10^{-20}\,{\#e_y}$~esu, $\#r_s= 10\,\#e_y$~nm%
\label{Fig:efieldy_centr}]%
{\includegraphics[width=10cm]{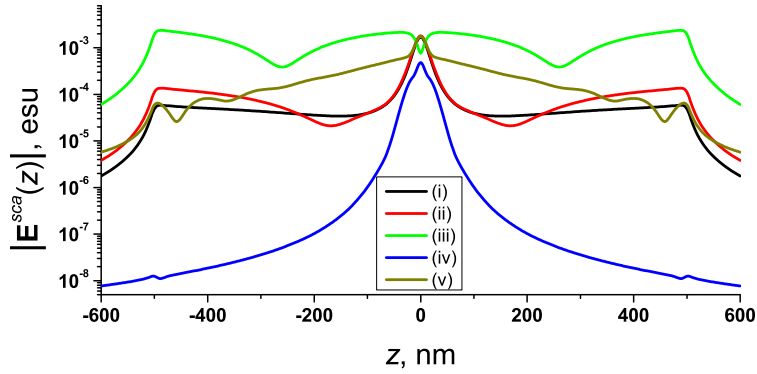}}%
\caption{\label{Fig:efield_centr}
Magnitude of the scattered electric field $\vert{\#E^{sca}}\vert$ computed
at $\#r= z\,\#e_z+ 10\,\#e_y$~nm, for $\vert{z}\vert \leq 0.6L$, when the SWNT and the illumination
conditions are the same as for Fig.~\ref{Fig:current_centr}.
The source electric dipole is oriented either (a) parallel or (b) normal to the axis of the SWNT.
}
\end{figure}

\begin{figure}[h]
\centering%
\includegraphics[width=10cm]{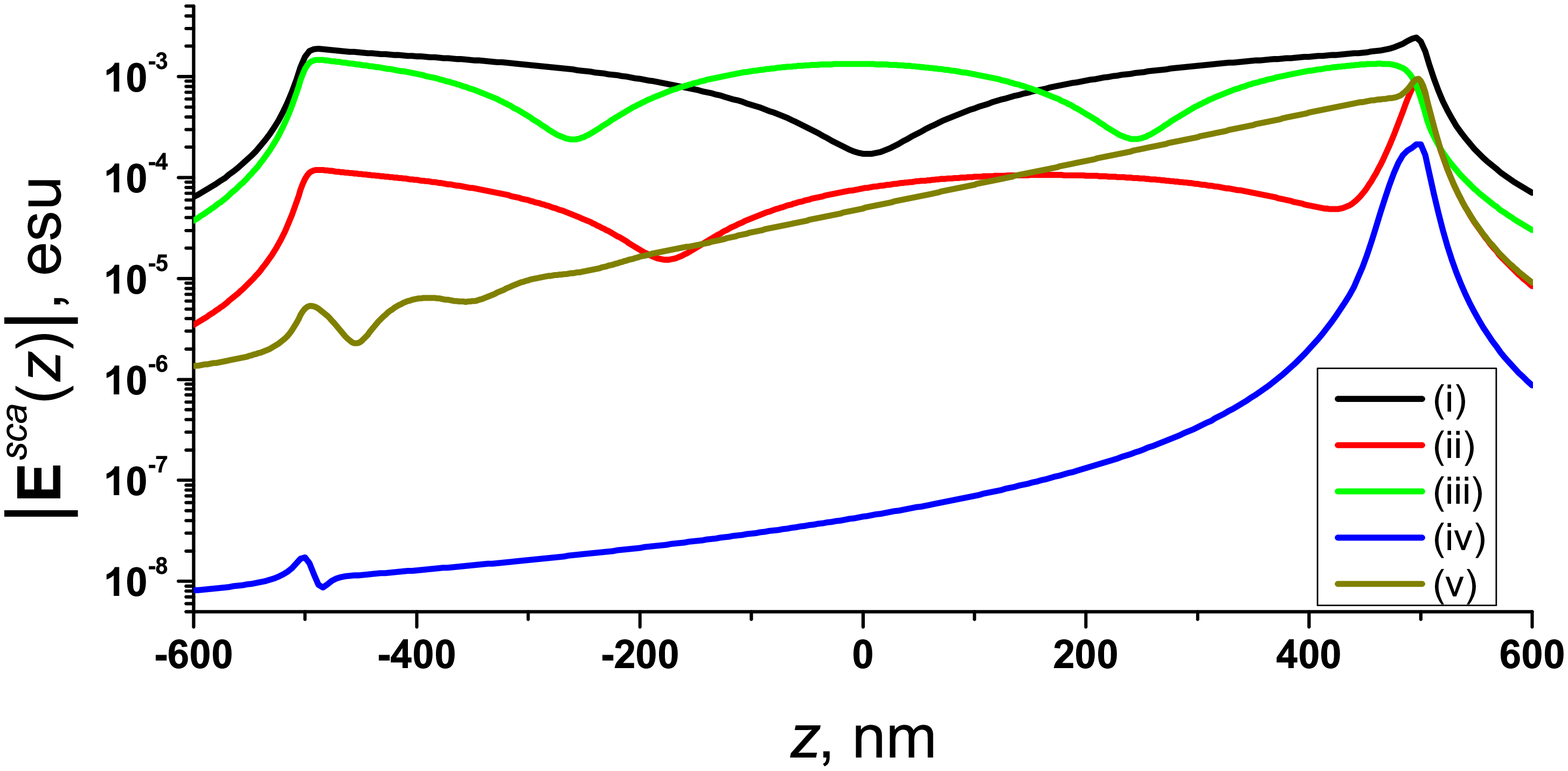}
\caption{\label{Fig:efield_edge}
Same as Fig.~\ref{Fig:efield_centr}, except that the irradiation
conditions are those for Fig.~\ref{Fig:current_edge} ($\#p_0=10^{-20}\,{\#e_z}$~esu
and $\#r_s= 510\,\#e_z$~nm).
}
\end{figure}

The scattered electric field in the vicinity of the SWNT was calculated using Eq.~(\ref{Esca}) for all five cases, and two locations and
two orientations (as appropriate)
of the source electric dipole. Figures~\ref{Fig:efield_centr} and \ref{Fig:efield_edge}
show  $\vert \#E^{sca}(\#r)\vert$ for $\#r = z\,\#e_z+10\,\#e_y$~nm, $\vert{z}\vert\leq
0.6L$.

Substantial enhancement of the maximum value  with increasing frequency is
not evident in Figs.~\ref{Fig:efield_centr} and \ref{Fig:efield_edge}. However, those figures do offer evidence of the effect
of dipole orientation on the scattered electric field. The plots in Figs.~\ref{Fig:efield_centr} and \ref{Fig:efield_edge} demonstrate that there are two spatial regions wherein the scattered field is localized:  (i) close to
the source electric dipole and (ii) near the edges of the SWNT. 
The localization of the field scattered by the SWNT in the region close to
source  electric dipole could be used in SNOM for the excitation of strongly localized electric fields caused by the spontaneous decay of an emitter placed in the vicinity of an SWNT.

\subsection{Radiation patterns of the dipole-SWNT system }
\label{Ch:Pattern}

Let us now consider the electromagnetic field in the far zone, with direct contribution from the source electric dipole and indirect contribution from the scattered field due to the presence of the SWNT. For this purpose, we chose a spherical coordinate
system $(r,\theta,\phi)$
with origin located at the centroid of the SWNT. We define
the joint radiation pattern
\begin{equation}
\#f(\#e_r) = \lim_{kr\to \infty}\,r\, e^{-ikr}\left[\#E^{{sca}}(r\#e_r)+\#E^{{inc}}(r\#e_r)\right]\,
\end{equation}
at distances far from the dipole-SWNT system.

 Equation~(\ref{Esca}) yields
\begin{equation} \label{Eq21}
\lim_{kr\to \infty}\,r\, e^{-ikr}\,\#E^{{sca}}(r\#e_r) = -\#e_{\theta}\frac{i2\pi{R_{cn}}\omega\sin\theta}{c^2}  \,
\int\limits_{-0.5L}^{0.5L} e^{-i k z \cos\theta} J^{eq}(z)dz\,,
\end{equation}
which does not depend on $\phi$,
whereas Eq.~(\ref{Eq:Field_inc1}) yields
\begin{equation}
\label{Eq22}
\lim_{kr\to \infty}\,r\, e^{-ikr}\,\#E^{{inc}}(r\#e_r) = k^2 e^{-ik\#e_r\cdot\#r_s}
(\#e_\theta\#e_\theta +\#e_\phi\#e_\phi)\cdot\#p_0\,.
\end{equation}
When $\#p_0\parallel \#e_z$, Eq.~(\ref{Eq22}) simplifies to
\begin{equation}
\label{Eq22z}
\lim_{kr\to \infty}\,r\, e^{-ikr}\,\#E^{{inc}}(r\#e_r) = -{k^2p_0} e^{-ik\#e_r\cdot\#r_s}\#e_\theta\,{\sin\theta}\,;
\end{equation}
when $\#p_0\parallel \#e_y$, we get
\begin{equation}
\label{Eq22y}
\lim_{kr\to \infty}\,r\, e^{-ikr}\,\#E^{{inc}}(r\#e_r) = {k^2p_0} e^{-ik\#e_r\cdot\#r_s}
\left(\#e_\theta\,{\cos\theta\sin\phi}+\#e_\phi\,\cos\phi\right)\,.
\end{equation}

Let us examine the function $\#f(\#e_r)$ when the source electric dipole is oriented
parallel to the SWNT axis and is located on the SWNT axis (i.e., $\#r_s=z_s\#e_z$); then,
\begin{equation}
\#f(\#e_r)= -\#e_{\theta}\,{k^2p_0}{\sin\theta}
\left[
e^{-ikz_s\cos\theta}+
\frac{i2\pi{R_{cn}}}{\omega{p_0}}  \,
\int\limits_{-0.5L}^{0.5L} e^{-i k z \cos\theta} J^{eq}(z)dz
  \right] \label{Eq:RadPatternZ}
\end{equation}
is independent of $\phi$. Displayed in Fig.~\ref{Fig:ScatPat} are polar plots of
the normalized joint radiation pattern
\begin{equation} \label{Eq:ScatPatNorm}
\gamma(\theta) = {\sin\theta}
\left\vert
e^{-ikz_s\cos\theta}+
\frac{i2\pi{R_{cn}}}{\omega{p_0}}  \,
\int\limits_{-0.5L}^{0.5L} e^{-i k z \cos\theta} J^{eq}(z)dz
  \right\vert
\end{equation}
in any plane to which the $z$ axis is tangential---for $z_s=0$ and $z_s=501$~nm,
and for the five different cases (i)--(v) delineated in Sec.~\ref{Ch:Numerical}.

\begin{figure}[h]
\centering%
\includegraphics[width=8.0cm]{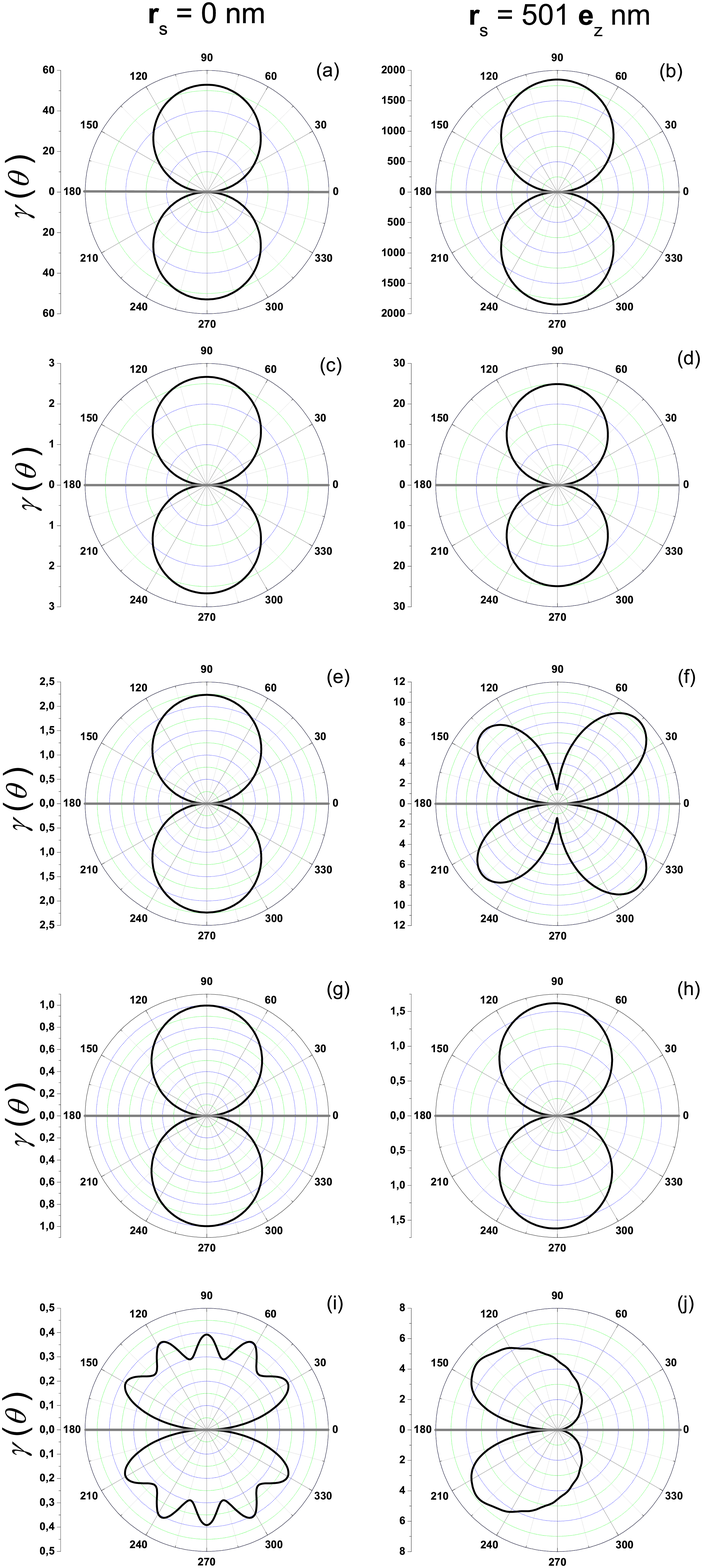}
\caption{\label{Fig:ScatPat}
Normalized joint radiation pattern $\gamma(\theta)$   for (left) $\#r_s=\#0$ and
(right) $\#r_s=501\,\#e_z$~nm, when $\#p_0=10^{-20}\,\#e_z$~esu. From top
to bottom:  (a, b)  case (i); (c, d)  case (ii); (e, f) case (iii); (g, h)   case (iv); and (i, j)   case (v). 
These calculations were made for a (15,0) SWNT (a-h) and (18,0) SWNT (i,j).
}
\end{figure}

In general, the plots of the normalized joint radiation pattern $\gamma(\theta)$ presented in Fig.~\ref{Fig:ScatPat} for a (15,0) and (18,0) SWNT do not have the $\sin\theta$ form expected of radiation from
point electric dipoles. The condition for $\gamma(\theta)$  having the dipolar form---as in
Figs.~\ref{Fig:ScatPat}(a)-(e)---is $kL\ll1$, which means that the second term on the right side of
Eq.~(\ref{Eq:ScatPatNorm}) can be ignored. But that is not enough. Indeed, in case (iii) for the source electric dipole placed near an edge of the SWNT edge, Fig.~\ref{Fig:ScatPat}(f) 
does not have the dipolar form even though   the condition  $k L \ll 1$ obviously holds true. Let us study this case in more detail.

Shown in Fig.~\ref{Fig:current_arg} the induced surface current density on a 
 (15,0) SWNT at 5.2~THz  when $\#p_0 = 10^{-20}\, \#e_z$~esu and
$\mathbf{r}_s = 501 \,\#e_z$~nm. Whereas the magnitude of $J^{eq}(z)$ is
symmetric with respect to the center plane $z=0$, the argument is
asymmetric. The integral on the right side of 
Eq.~(\ref{Eq:ScatPatNorm}) is therefore almost null valued for $\theta=\pi/2$
and $\gamma(\pi/2)$ is small.
However, for $\theta \neq \pi/2$ the phase of the integrand changes due to the 
presence of $e^{-i k z \cos\theta}$, and the joint radiation pattern is enhanced.  As
the right side of Eq.~(\ref{Eq:ScatPatNorm}) contains terms proportional
to   $\cos\theta$  and $\sin\theta$, radiation lobes centered about $\vert\sin\theta\vert=
\vert\cos\theta\vert=1/\sqrt{2}$ appear.

\begin{figure}[ht!]
\begin{center}
\includegraphics[width=10cm]{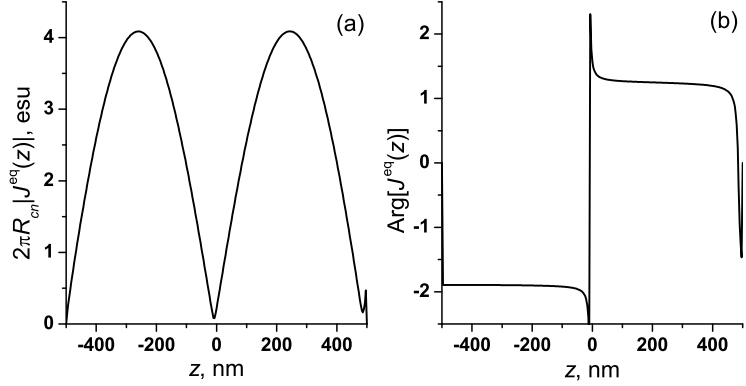}
\caption{(a) Magnitude  and (b) argument  of the surface current density induced on a (15,0) SWNT
by a source electric dipole of moment  $\#p_0 = 10^{-20}\, \#e_z$~esu located at
$\mathbf{r}_s = 501\, \#e_z$~nm. The frequency is 5.2~THz.}
 \label{Fig:current_arg}
\end{center}
\end{figure}

At sufficiently high frequencies, the condition $k L \ll 1$ does not hold.
However, we still observe the dipolar radiation pattern in Figs.~\ref{Fig:ScatPat}(g,h)
for case (iv). When the source electric dipole is placed at the 
centroid of the SWNT, as for Fig.~\ref{Fig:ScatPat}(g), the main contribution to the normalized joint radiation
pattern is from the first term on the right side of Eq.~(\ref{Eq:ScatPatNorm}). That term
is manifestly dipolar.
Scattering by the SWNT is small in the high-frequency regime, due to the strong damping of the surface wave, leading to ohmic absorption rather than reradiation. In contrast,
when the source electric dipole is placed near one edge of the SWNT, neither of
the two terms  on the right side of Eq.~(\ref{Eq:ScatPatNorm}) is dominant. But
the normalized joint radiation pattern in Fig.~\ref{Fig:ScatPat}(h) is still dipolar
because, as the induced current is confined largely to
the  edge close to the source electric
dipole, that edge acts like an additional electric dipole. Accordingly, we conclude
that the presence of an SWNT does not
lead to PL enhancement  at optical frequencies---in good agreement with the experimental results where strong luminescence quenching for Gd-Se quantum dots and polystyrene spheres in the presence
of SWNTs has been observed \cite{Mu}. 

Figures \ref{Fig:ScatPat}(i,j) indicate that the joint radiation pattern of the dipole-SWNT
system is very different from that of the dipole alone, in case (v). This happens because
the chosen frequency is a plasmon resonance frequency.

The normalized joint radiation patterns presented in Fig.~\ref{Fig:ScatPat} also demonstrate 
the crucial influence of the edges of the SWNT on PL enhancement. At all frequencies the far-field radiation intensity is much higher when the source electric dipole is placed
near an edge of SWNT (right column in Fig.~\ref{Fig:ScatPat}) than  at the centroid of the SWNT (left column in Fig.~\ref{Fig:ScatPat}). A strong enhancement by a factor  $\sim2000$ is predicted at the frequency of the  first geometrical resonance, i.e., case (i).


\subsection{Resolution with an SWNT tip for sSNOM}
\label{Ch:2D}

As stated in Sec.~\ref{intro}, the formalism presented in this paper can be
applied for sSNOM investigations of a PL object using an SWNT tip.  In order
to exemplify that assertion, we fixed a (15,0) SWNT 
of length $L=1$~$\mu$m on the $z$ axis with the
centroid of the SWNT serving as the origin of the coordinate system, varied
the location $\#r_s = x_s \,\#e_x+y_s\,\#e_y+505\,\#e_z$~nm
 of the source electric dipole of moment $p_0 = 10^{-20}$~esu radiating
at   2.6~THz frequency, and computed $\vert\mathbf{f}(\mathbf{e}_x)\vert/k^2 p_0$
as a function of $x_s$ and $y_s$ for both $\#p_0\parallel\#e_z$ and $\#p_0\parallel\#e_y$.
The resulting contour plots are presented in Fig. \ref{Fig:ScatMapZ}.

\begin{figure}[!h]
\centering%
\includegraphics[width=\textwidth]{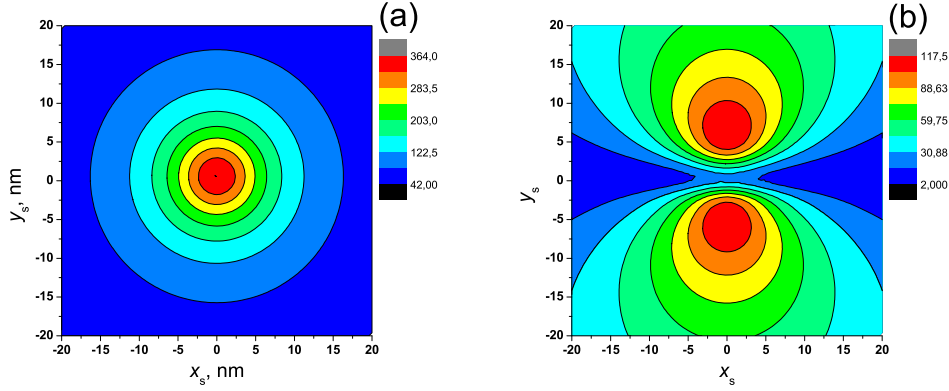}
\caption{\label{Fig:ScatMapZ} Contour plot of
$\vert\mathbf{f}(\mathbf{e}_x)\vert/k^2 p_0$ as a function of
$x_s$ and $y_s$, when a source electric
dipole radiating
at   2.6~THz frequency is located at
$\#r_s = x_s \,\#e_x+y_s\,\#e_y+505\,\#e_z$~nm, and
a (15,0) SWNT 
of length $L=1$~$\mu$m is affixed to the $z$ axis with the
centroid of the SWNT serving as the origin of the coordinate system. The dipole electric moment is (a) $\mathbf{p}_0 = 10^{-20}\, \mathbf{e}_z$~esu; (b) $\mathbf{p}_0 = 10^{-20}\, \mathbf{e}_y$~esu. 
}
\end{figure}

When the electric dipole is oriented along the SWNT axis, the contour plot of the joint radiation pattern in Fig.~\ref{Fig:ScatMapZ}a is a set of concentric circles centered on the SWNT axis. The jointly radiated field decreases in strength rapidly: within $20$~nm from the SWNT axis, there is a drop in magnitude by one order. When the electric dipole is oriented normal
the SWNT axis, the contour plot   in Fig.~\ref{Fig:ScatMapZ}b  shows two images arranged 
symmetrically with respect the plane formed by the SWNT axis and the direction normal to both the SWNT axis and the orientation of the electric dipole. The two images arise due to
 the weak coupling of $y$-oriented dipole with the SWNT when the distance between the two is small.  

In order to study the spatial resolution of the SWNT tip as a probe,
the calculations for  Fig.~\ref{Fig:ScatMapZ} were repeated but with
\textit{two} $z$-oriented electric dipoles, one placed at $\#r_{s1} = \#r_{c} + (x_s\#e_x  + y_s \#e_y )/2$
and the other at $\#r_{s2}=\#r_{c} - (x_s\#e_x  + y_s \#e_y)/2$, where $\mathbf{r}_c =x_c \,\#e_x+y_c\,\#e_y+505\,\#e_z$~nm is the radius-vector of the two dipoles system geometrical center, $x_s$ and $y_s$ are the interdipole separations along the $x$ and $y$-axis respectively. In the contour plots presented in 
Fig.~\ref{Fig:ScatMapZ2}, the two  electric dipoles cannot be resolved
when the inter-dipole separation is $10$~nm. However, resolution is
possible when that separation is $20$~nm, leading us to the conclusion that spatial resolution of the SWNT-based probe   could be between 10 and 20~nm.

\begin{figure}[h]
\centering%
\includegraphics[width=\textwidth]{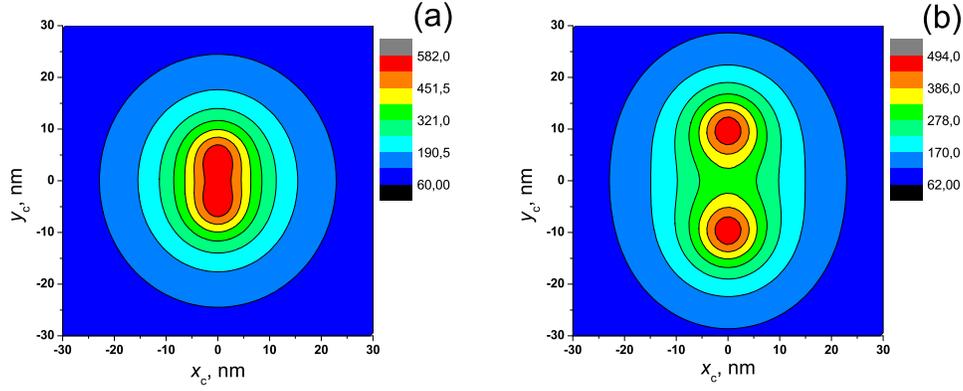}
\caption{\label{Fig:ScatMapZ2}
Contour plots of
$\vert\mathbf{f}(\mathbf{e}_x)\vert/k^2 p_0$ as a function of
$x_c$ and $y_c$, when two identical electric
dipoles of moment $\mathbf{p}_0 = 10^{-20}\, \mathbf{e}_z$~esu radiating
at   2.6~THz frequency are located at
$\#r_{s1} = x_c \#e_x  + (y_c + y_s/2)\#e_y + 505 \#e_z$~nm and $\#r_{s2} = x_c \#e_x  + (y_c - y_s/2)\#e_y + 505 \#e_z $~nm,
and
a (15,0) SWNT 
of length $L=1$~$\mu$m is affixed to the $z$ axis with the
centroid of the SWNT serving as the origin of the coordinate system.
Inter-dipole separation is (a) $y_s = 10$~nm and (b) $y_s = 20$~nm.
}
\end{figure}

\section{Concluding remarks}
\label{Ch:Conclusion}

Following a standard procedure, we formulated a Fredholm integral equation for the
surface current density induced on a metallic SWNT irradiated by the electromagnetic field of
an arbitrary source located outside the SWNT. The integral equation
was solved numerically and then used to compute the scattered field.
Though we chose the source to be
a point electric dipole for numerical work, the technique can be used for other sources
such as electrically small loop antennas \cite{LIDH82a} and even extended sources
such as aperture antennas \cite{Wood}.

The relative location and orientation of the source electric dipole influence
the profile of the current induced in the SWNT as {well as } the scattered electric field
in the vicinity of the SWNT. This {effect } can be accentuated by the {resonant }
excitation of a surface wave on the SWNT.  {We proved this by investigating the frequency dependence of the induced current.}
 Strong spatial localization of scattered electric field near one or both edges of the SWNT,
 particularly under nonresonant conditions, should promote the adoption of SWNT tips
 for
 scanning near-field optical microscopy.

Carbon nanotubes are strongly nonlinear, the nonlinearity becoming substantial when the incident power
density is on the order of
$10^{10}$~W~cm$^{-2}$ or larger \cite{Stanciu_02}. For a gaussian pulse, this power density corresponds to a
peak electric field on the order of $10^{7}$~V~cm$^{-1}$, which is three orders of magnitude lower than the atomic field.  This value of the electric field strength is of the same order of magnitude as the electric field strength at a distance of 
$1$~nm from the electric dipole---see Eq. (\ref{Eq:Field_inc1})---with dipole moment  $p_0 = 30$~D $= 3\times 10^{-17}$~esu which is the typical value of the  dipole moment of a quantum dot \cite{Silverman}. Thus, though the power density in the near-field region can be
quite  low, the electric field strength   can be  high enough such that nonlinear effects become
significant. Thus, a
 comprehensive description of the scattering of a near field by an SWNT
will lead to the development of \textit{near-field nonlinear optics} of SWNTs.

\section*{Acknowledgments}
This research was partially supported by
the International Bureau BMBF
(Germany) under project BLR 08/001, the Belarus Republican
Foundation for Fundamental Research (BRFFR) under project F09MC-009 and EU FP7
CACOMEL project FP7-247007.
AL thanks the Charles Godfrey Binder Endowment at Penn State for partial financial support of his research activities.
The work of AMN was supported by the BRFFR young scientists grant F09M-071.

\vspace{2ex}
\begin{wrapfigure}{l}{1.8cm}
\includegraphics[width=1.8cm]{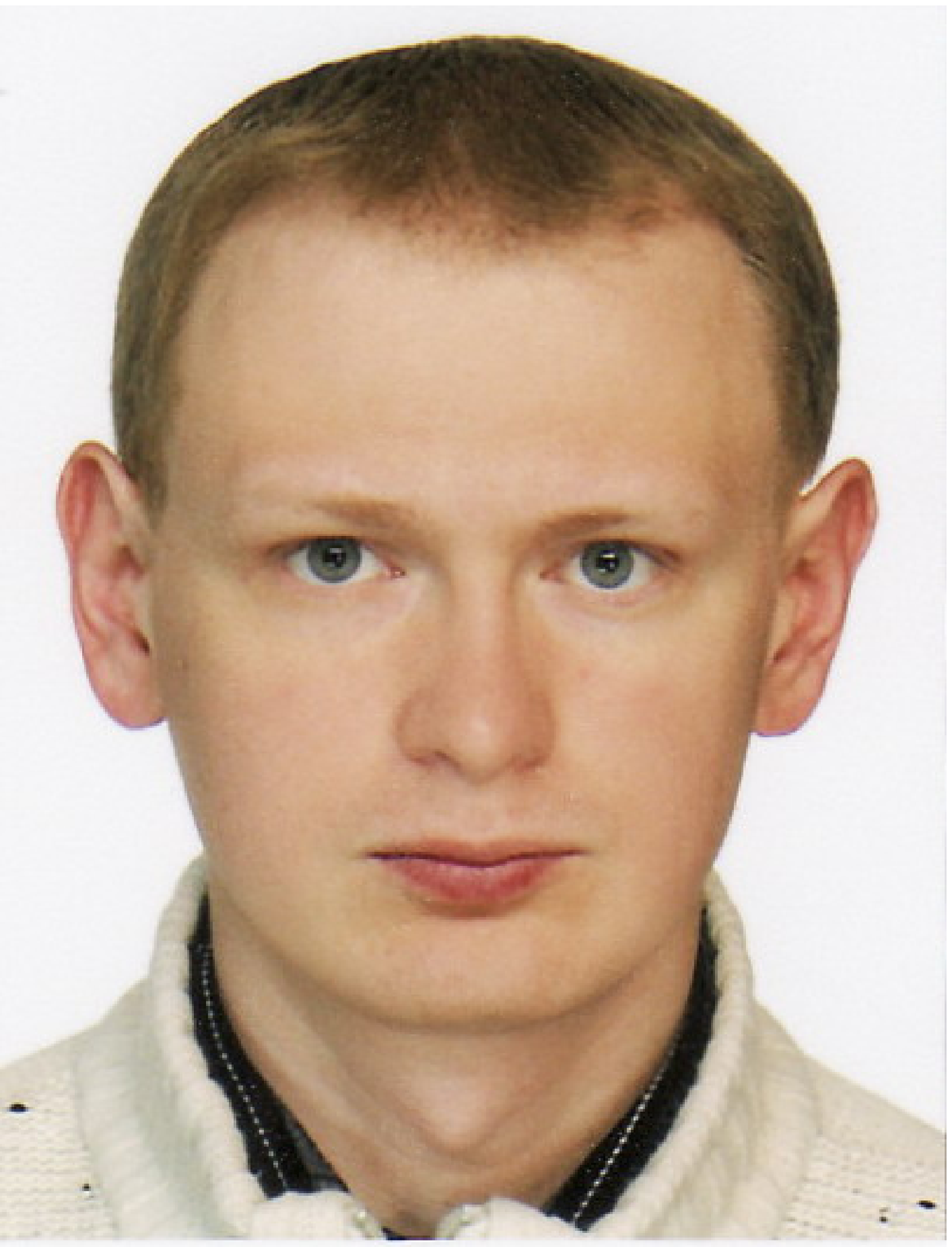}
\end{wrapfigure}
\noindent{\bf Andrei M. Nemilentsau} received his M.S. in theoretical physics in 2004 from Belarus State University (BSU), Minsk, Belarus, and his Ph.D. in theoretical physics in 2009 from the Institute of Physics, Belarus National Academy of Sciences, Minsk, Belarus. Currently he is working as a scientific researcher in the Laboratory of Electrodynamics of Nonhomogeneous Media at the Institute for Nuclear Problems, BSU. His current research interests are the electromagnetic processes in nanostructures. Dr. Nemilentsau is a recipient of  the World Federation of Scientists Scholarship for Young Scientists (2003) and the INTAS Young Scientist Fellowship for Ph.D. Students (2006).

\vspace{2ex}
\begin{wrapfigure}{l}{1.8cm}
\includegraphics[width=1.8cm]{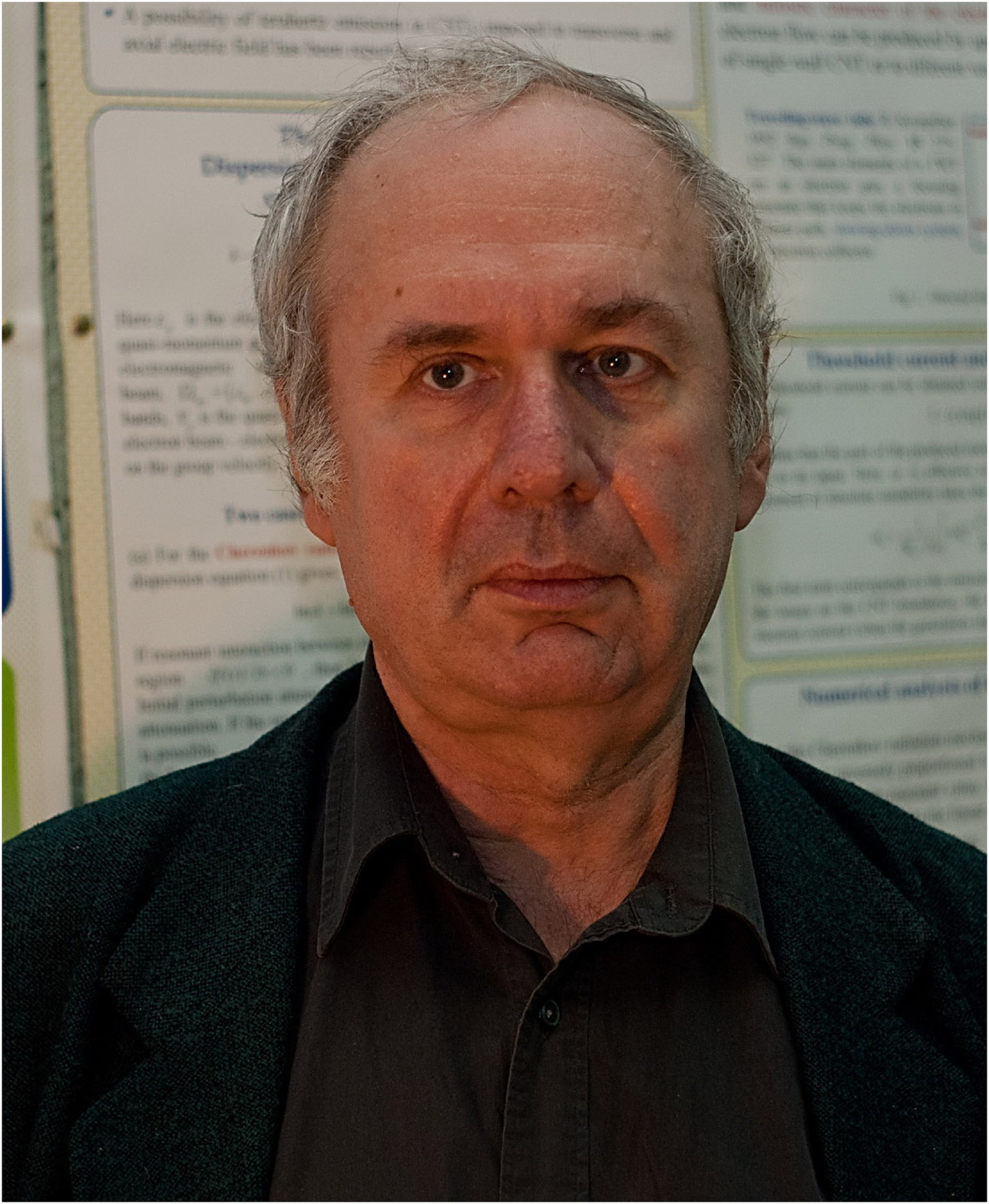}
\end{wrapfigure}
\noindent{\bf Gregory Ya. Slepyan} received his M.S. in radioengineering in 1974 from Minsk Institute of Radioengineering, Minsk, Belarus; his Ph.D. in radiophysics in 1979 from Belarus State University (BSU), Minsk, Belarus; and his D.Sc. in radiophysics in 1988 from Kharkov State University, Kharkov, Ukraine. Since 1992 he has been working as a principal researcher of the Laboratory of Electrodynamics of Nonhomogeneous Media at the Institute for Nuclear Problems, BSU. He has authored or coauthored 2 books, 4 collective monographs, and more than 200 conference and journal papers. He is the Member of Editorial board of the journal \textit{Electromagnetics}. His current research interests are high-power microwave vacuum electronics, antennas and microwave circuits, mathematical theory of diffraction, nonlinear oscillations and waves, nanophotonics and nano-optics (including electrodynamics of carbon nanotubes and quantum dots).

\vspace{2ex}
\begin{wrapfigure}{l}{1.8cm}
\includegraphics[width=1.8cm]{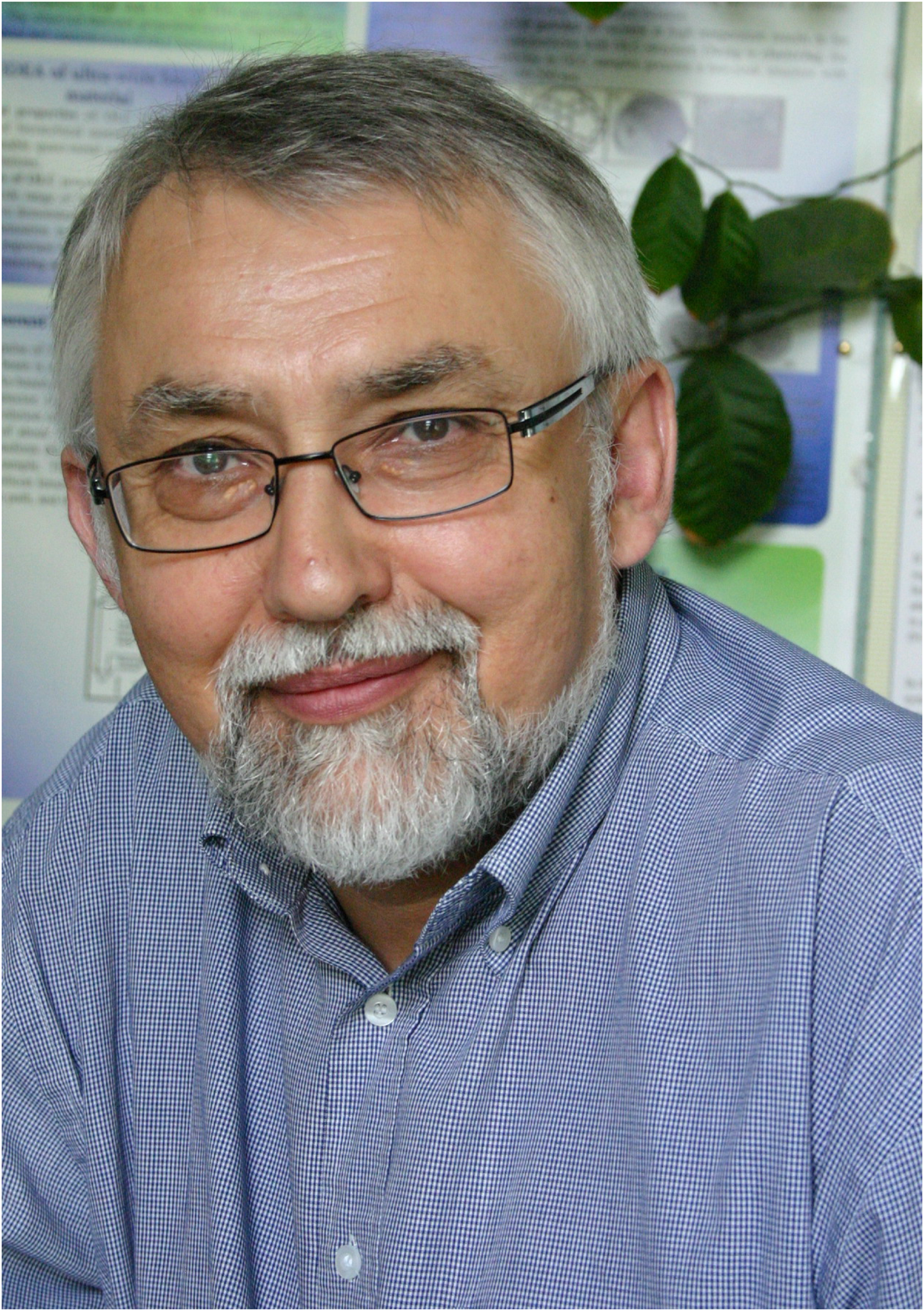}
\end{wrapfigure}
\noindent{\bf Sergei A. Maksimenko} received his M.S. in physics of heat and mass transfer in 1976 and his Ph.D. in theoretical physics in 1988, both from Belarus State University (BSU), Minsk, Belarus, and his D.Sc. in theoretical physics in 1996 from the Institute of Physics, Belarus National Academy of Sciences, Minsk, Belarus. Since 1992 he has been working as head of the Laboratory of Electrodynamics of Nonhomogeneous Media at the Institute for Nuclear Problems (INP), BSU. He also teaches at the BSU physics department. He was the deputy director of INP from 1997 to 2000,  and a deputy vice-president  of BSU from 2000 to 2005.  He has authored or coauthored more than 150 conference and journal papers. In 2003, 2004 and 2006 years he co-chaired conferences as parts of SPIE's 48th and 49th Annual Meetings, and SPIE's Optics and Photonics. He is an associate editor of \textit{Journal of Nanophotonics} and an SPIE Fellow. His current research interests are electromagnetic wave theory and electromagnetic processes in quasi-one- and zero-dimensional nanostructures in condensed matter.

\vspace{2ex}
\begin{wrapfigure}{l}{1.8cm}
\includegraphics[width=1.8cm]{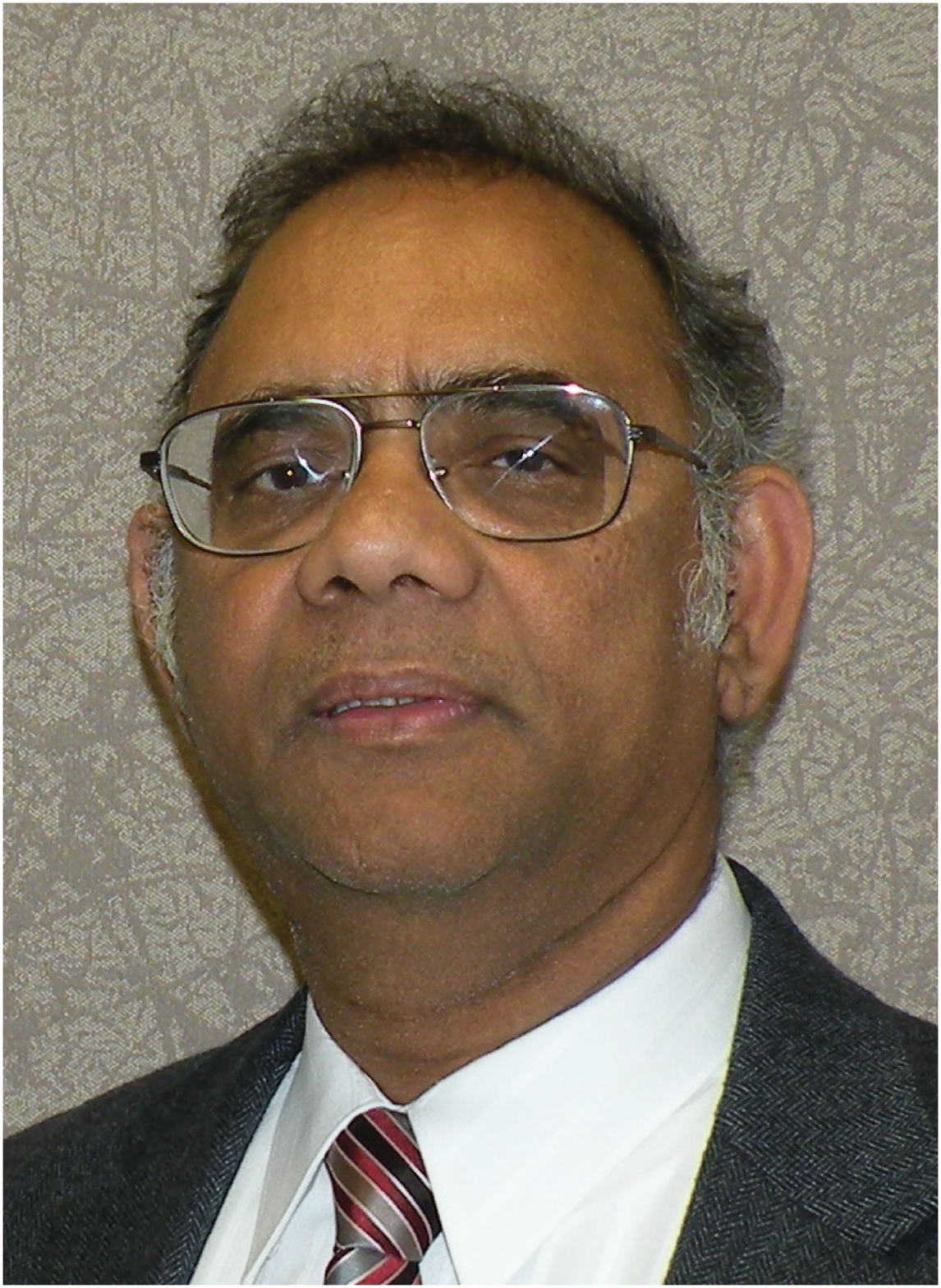}
\end{wrapfigure}
\noindent {\bf Akhlesh Lakhtakia} received degrees from the Banaras
Hindu University (B.Tech. \& D.Sc.) and the University of Utah (M.S.
\& Ph.D.), in electronics engineering and electrical engineering,
respectively. He is the Charles Godfrey Binder (Endowed) Professor
of Engineering Science and Mech\-anics at the Pennsylvania State
University.   He is a Fellow of SPIE, Optical Society of America, 
American Association for the Advancement of Science, and Institute of
Physics (UK). He is presently the editor-in-chief of \textit{Journal
of Nanophotonics}. His current research interests include nanotechnology,
plasmonics, complex materials, metamaterials, and sculptured thin
films.

\vspace{2ex}
\begin{wrapfigure}{l}{1.8cm}
\includegraphics[width=1.8cm]{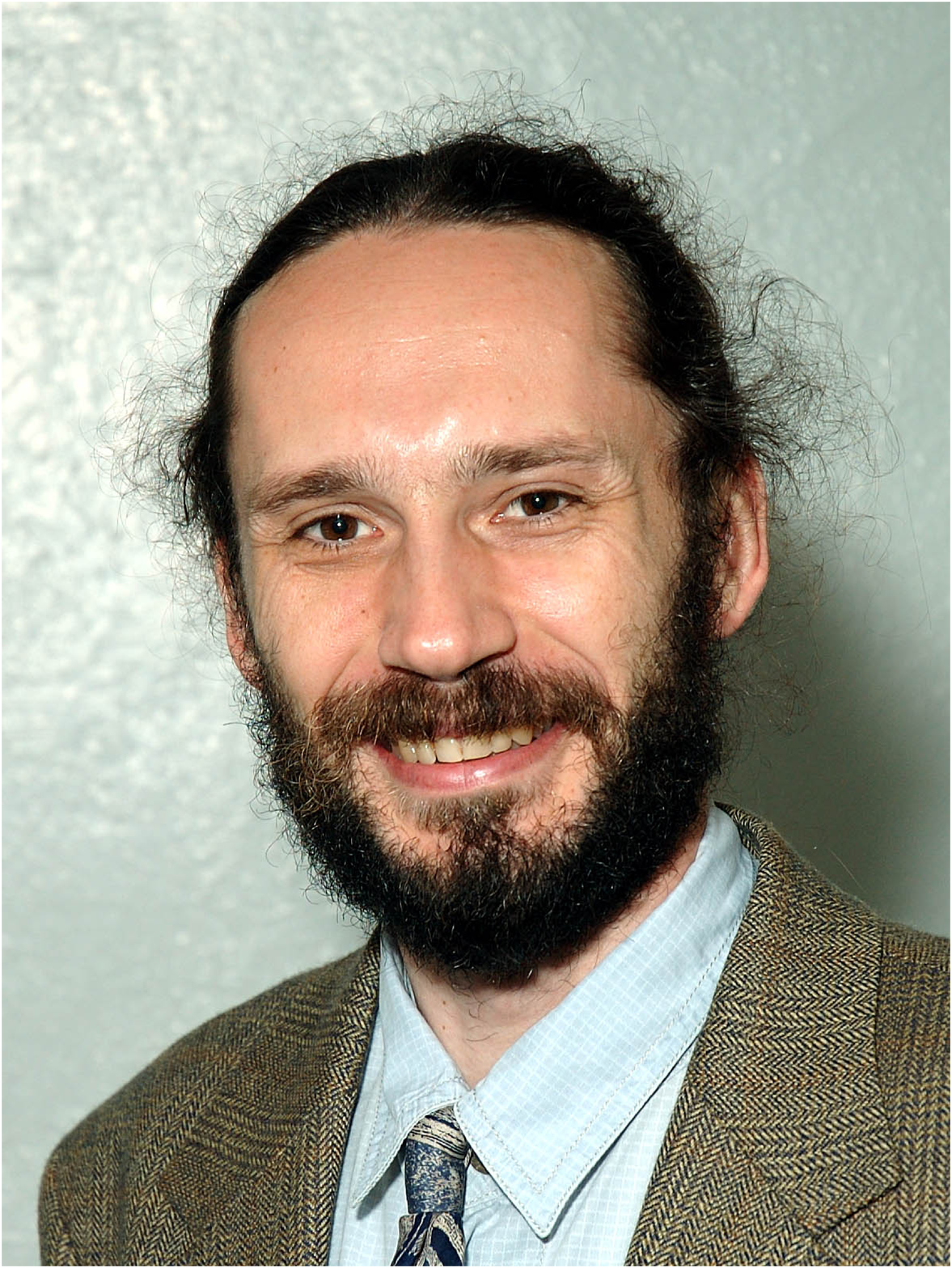}
\end{wrapfigure}
\noindent{\bf Slava V. Rotkin} is the Frank J. Feigl Junior Faculty Scholar and Assistant Professor
of Physics at Lehigh University, Bethlehem, PA. He received his M.Sc. (Summa Cum Laude) in 1994 from
the Electro-technical University and his Ph.D. in 1997 from Ioffe Institute (both in St. Petersburg, Russia).
Dr. Rotkin is a recipient of scientific awards, including: Libsch Early Career Research Award (2007),
Feigl Scholarship (2004),  Beckman Fellowship (2000),  Royal Swedish Academy of Sciences Fellowship (1995),
and  President's Grant for Young Scientists of Russia (1994).

\end{document}